\documentclass{iopart}
\usepackage{fullpage}
\usepackage{amstext}
\usepackage{iopams}
\pdfoutput=1

\usepackage{graphicx}
\usepackage[pdftex,colorlinks=true,urlcolor=purple,filecolor=purple,citecolor=purple,hyperfootnotes=true]{hyperref}
\usepackage[latin1]{inputenc}
\usepackage{xcolor}
\usepackage{ulem}

\newcommand{\bol}[1]{{\boldsymbol #1}}
\newcommand{\eqref}[1]{({\ref{#1}})}

\renewcommand{\H}{{\cal H}}
\newcommand{\eps}{\varepsilon}

\renewcommand{\d}{{\rm d}}
\newcommand{\dt}{{\rm d}t}
\newcommand{\ee}{{\rm e}}
\newcommand{\ddt}{\frac{\d}{\d t}}

\newcommand{\ddp}[2]{\frac{\partial {#1}}{\partial {#2}}}

\begin{document}

\author{Tanguy Laffargue$^{1}$, Khanh-Dang Nguyen Thu Lam$^{2}$, Jorge
  Kurchan$^{2}$, Julien Tailleur$^{1}$}

\address{\ $^1$ Univ Paris Diderot, Sorbonne Paris {Cit\'e}, MSC, UMR 7057 CNRS, F75205 Paris, France}

\address{\ $^2$ Laboratoire PMMH (UMR 7636 CNRS, ESPCI, P6, P7),
         10 rue  Vauquelin, 75231 Paris cedex 05, France}

\title[Large deviations of Lyapunov exponents]{Large deviations of Lyapunov exponents}

\date{\today}

\begin{abstract}
Generic dynamical systems have `typical' Lyapunov exponents, measuring
the sensitivity to small perturbations of almost all trajectories.  A
generic system has also trajectories with exceptional values of the
exponents, corresponding to unusually stable or chaotic situations.
From a more mathematical point of view, large deviations of Lyapunov
exponents characterize phase-space topological structures such as
separatrices, homoclinic trajectories and degenerate tori. Numerically
sampling such large deviations using the \textit{Lyapunov Weighted
  Dynamics} allows one to pinpoint, for example, stable configurations
in celestial mechanics or collections of interacting chaotic
`breathers' in nonlinear media. Furthermore, we show that this
numerical method allows one to compute the topological pressure of
extended dynamical systems, a central quantity in the Thermodynamic of
Trajectories of Ruelle.
\end{abstract}


\maketitle

\section{Introduction}

\vspace{.5cm}

In many physical systems, the typical situations are not the most
interesting, and one is led to look for rare trajectories. For
example, of all possible initial conditions of eight planets, only a
small number lead to a stability comparable to the one of the solar
system \cite{L1,MH1,MH2}. Exceptional configurations such as
resonances and separatrices play an important role in this kind of
system \cite{L1}.  Similarly, a study of the transport properties of
almost-integrable systems requires the knowledge of chaotic layers
that are extremely thin, because they are the structures responsible for
the global diffusion mechanism~\cite{Arnold64,Froeschle,Seibert}.  Apart from being
rare, often these relevant trajectories turn out to be unstable, a
further source of difficulty.  For example, unstable soliton and
breather modes may be important factors in the transport of energy of
Bose-Einstein condensates, and even macromolecules
\cite{Ruffo1,TS}. Similarly, the phenomenon of intermittency, which
has a profound effect in turbulent systems, is believed to be
generated by localized spatial structures that appear and disappear in
the flow {\cite{instantons}.}
 
Each physical situation will have one or more quantities whose large
deviations are relevant. For example, for a rare `rogue' wave, a
quantity that is certainly interesting is its energy or its height,
for a traffic problem the actual flow, and for a planetary orbit its
eccentricity or inclination angle. Large deviations of Lyapunov
exponents are a particular and important case, as has been recognized
for years \cite{Ruelle,paladin,Grassberger0}, which still attracts
lots of attention from the dynamical system community
nowadays~\cite{Vallejo2008,Yanagita2009,Kuptsov2011,Vallejos2012}.

\vspace{.25cm}

{\bf Large Deviations}

\vspace{.25cm}

For a dynamical system defined by trajectories of $n$-dimensional
variables ${\bf x}(t)$, consider an observable $A({\bf x})$. We wish
to study the trajectories such that, for example, $\int_0^\tau dt \;
A({\bf x}) = \tau A_o$. In particular, their total probability:
\begin{equation}
P(A_o,\tau)= \Big\langle \delta\Big(\int_0^\tau  dt \; A({\bf x}) -\tau A_o\Big)\Big\rangle
\end{equation}
A more convenient object to study is the Laplace transform of
$P(A_o,\tau)$
\begin{equation}
{Z(\alpha,\tau) = \left\langle e^{\alpha \tau A_o } \right\rangle= \int dA_o e^{\alpha \tau A_o} P(A_o,\tau) }
\end{equation}
Note that $\alpha$ is to $A_o$ what the inverse  temperature is to the energy
(density) in statistical mechanics: $Z(\alpha,\tau)$ is nothing but a partition function in the space
of trajectories and the large deviation formalism thus allows one to
extend the static formalism of statistical physics to dynamical
observables~\cite{Ruelle,beck}. It is thus natural to define
\begin{equation}
  \mu(\alpha,\tau) = \frac 1 \tau \log Z(\alpha,\tau)
\end{equation}
which plays the role of a free energy and
\begin{equation}
  A_o(\alpha,\tau)= \mu'(\alpha,\tau) = \frac{\langle A_o e^{\alpha \tau
      A_o} \rangle}{\langle e^{\alpha \tau
      A_o} \rangle}
\end{equation}
which plays the role of an order parameter, distinguishing between
various classes of trajectories with markedly different values of
$A_o$.
\if{  Often more interesting is the value other observables $B({\bf
  x})$ take when the system is conditioned as above:
\begin{equation}
\langle B\rangle_{A_o} = 
\frac{\left\langle \delta\left(\int_0^\tau  dt \; A({\bf x}) -\tau A_o\right)\; B({\bf x,t}) \right\rangle}{\left\langle \delta\left(\int_0^\tau  dt \; A({\bf x}) -\tau A_o\right)\right\rangle}
\end{equation}}\fi

\vspace{.25cm}

{\bf Lyapunov exponents and their large deviations}

\vspace{.25cm}

Consider two nearby points ${\bf x}_1(t=0)$ and ${\bf x}_2(t=0)$ and their subsequent trajectories  ${\bf x}_1(t)$ and ${\bf x}_2(t)$ in a dynamical system.  The distance ${\bf u}_1(t) = {\bf x}_1(t)-{\bf x}_2(t)$ at long times will typically grow as
\begin{equation}
\ln |{\bf x}_1(t)-{\bf x}_2(t)| \sim t \lambda_1
\end{equation}
The quantity $\lambda_1 \equiv \Lambda_1 $ is a measure of the
sensitivity to the initial conditions.  Similarly, we may consider
three non-colinear nearby points, and the two differences $\bol u_1(t)={\bf
  x}_1(t)-{\bf x}_2(t)$ and $\bol u_2(t)={\bf x}_2(t)-{\bf x}_3(t)$. The
lengths of $u_1(t)$ and $u_2(t)$ will grow as above, but we may ask
about the area $|{\bf u}_1 \wedge {\bf u}_2|$ of the parallelogram
defined by them. This defines a new exponent:
\begin{equation}
\ln |{\bf u}_1 \wedge {\bf u}_2| \sim t \Lambda_2 = t (\lambda_1+\lambda_2)
\end{equation}
In general, for generic points and  $p$ vectors $\{{\bf u}_1, ...,{\bf u}_p\}$, the $p$-volume grows as:
  \begin{equation}
    {\ln |{\bf u}_1 \wedge {\bf u}_2\wedge ...\wedge {\bf u}_p| = \frac{1}{2} \ln |\det \{{\bf u}_a . {\bf u}_b\} | \sim t \Lambda_p =  t (\lambda_1+\lambda_2+...+\lambda_p)}
\end{equation}
The $\lambda_a$ are the (finite-time) Lyapunov exponents~\footnote{{To
  maintain the introduction as free of mathematical technicalities as
  possible, we reserve more precise definitions of (finite-time)
  Lyapunov exponents for section~\ref{sec:method}}}, \if{, which may be
  also expressed by saying that the eigenvalues of the matrix with
  dimension equal to the full dimension of the space $\ln U= \ln
  \{{\bf u}_a . {\bf u}_b\} $ go as ${\frac{1}{2} t\lambda_a}$.}\fi
whose large deviations will be the focus of this article. This means
that we shall consider intervals of time $\tau$, and the role of
{$A_0$} above will be played by, for example, {$A_0=
  \lambda_1$}. {We can then classify trajectories according to
  the value of their first Lyapunov exponent $\lambda_1^o$ and
  construct the probability distribution:}
\begin{equation}
P(\lambda_1)= \left\langle \delta\left(\lambda_1 - \lambda_1^o\right)\right\rangle
\end{equation}
{and the corresponding moment- and cumulant-generating functions}
\begin{equation}
Z(\alpha,\tau) = \left\langle e^{\tau \alpha \lambda_1 }
\right\rangle\qquad \mu(\alpha,\tau)=\frac 1 \tau \log Z(\alpha,\tau)
\end{equation}
and their generalizations to higher Lyapunov exponents.
 
At this stage one may ask what is special about large deviations of
Lyapunov exponents, that this should merit a separate article.  There
are a number of reasons why we think this is so.  To begin with, it is
clear that Lyapunov exponents are a measure of chaoticity, and rare
trajectories having unusually low or unusually high values will
represent havens of stability (important for, e.g., planetary systems),
or chaotic bursts. We shall see several examples of this below.

The sum of the positive Lyapunov exponent is also related (via Pesin's
theorem~\cite{Ott,beck}) to the Kolmogorov-Sinai entropy: a space-time
quantity `counting' the number of \textit{typical trajectories}, a
useful measure of the geometry of the system.

Another important measure is the dimension of attractors. These are
related to the cumulative Lyapunov exponents $\Lambda_i$.  Consider
the value of $p$ such that $\Lambda_p>0$ and $\Lambda_{p+1}<0$.  The
dimension of the attractor has to be intermediate between {$p$ and
  $p+1$}, so that a volume element of the attractor advected by the
dynamics neither overflows nor contracts on it: the precise value is
encoded in the Kaplan-Yorke formula \cite{Ott,beck}.

\begin{figure}[h!]
\begin{center}
\includegraphics[width=10cm]{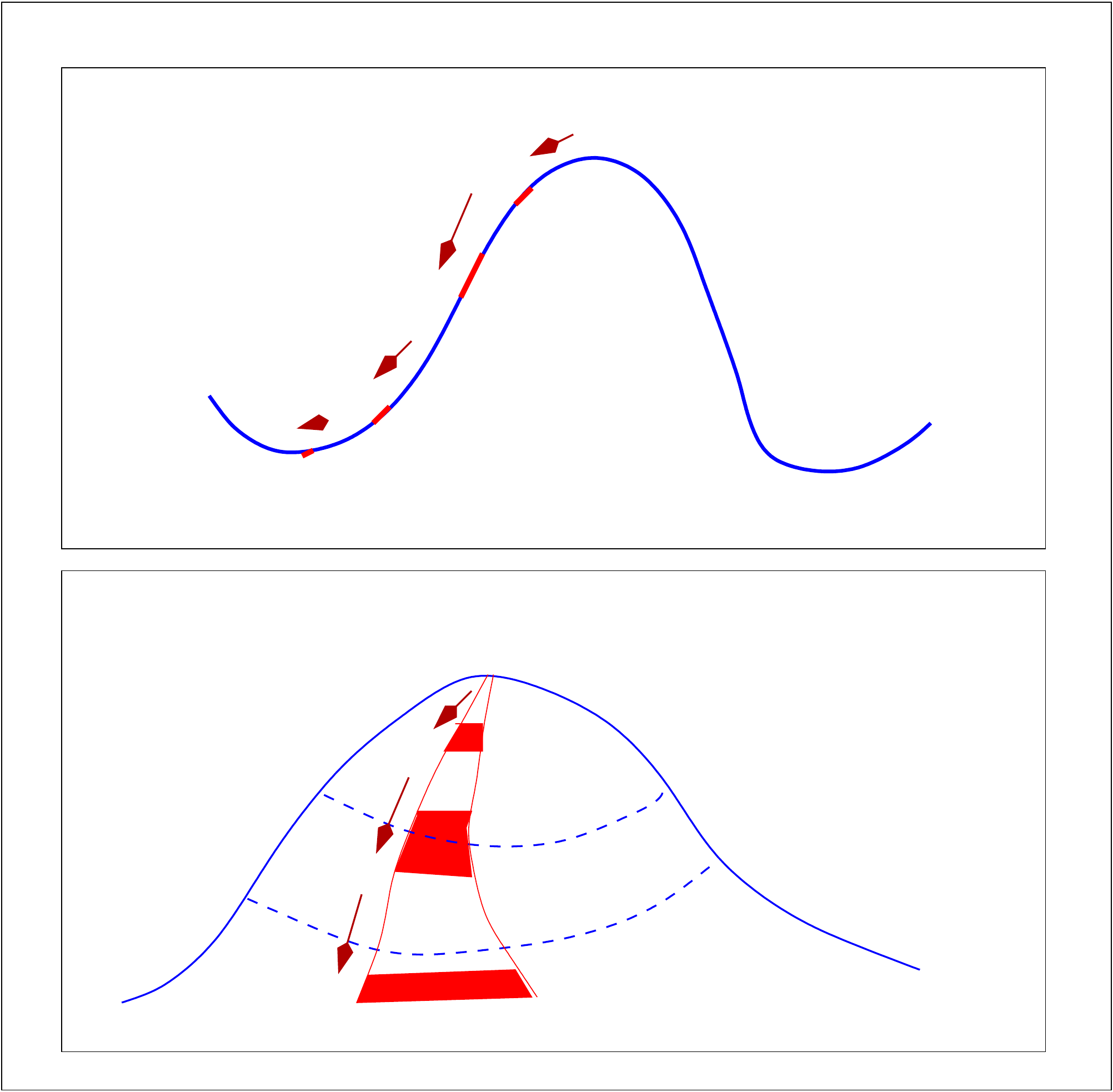}
\end{center}
\caption{Lyapunov exponents and topological structures. The volume
  contractions and expansions induced by evolution are exactly the
  instantaneous growth of the sums of the $p$ largest Lyapunov
  exponents. {\bf Top}: The weight $\exp(\lambda_1 t)$ stabilizes the
  unstable manifold emerging from critical points with one unstable
  direction. {\bf Bottom}: The weight $\exp[(\lambda_1+\lambda_2) t]$
  stabilizes the unstable manifold emerging from critical points with
  two unstable directions.}\label{fig:flow}
\end{figure}

{An even more specific property of Lyapunov large deviations is that it
allows us to pinpoint important topological features in the
dynamics. For instance, the family
\begin{equation}
Z_p(\alpha=1,\tau) = \langle e^{\tau \Lambda_p} \rangle = \langle e^{\tau \sum_{a=1}^p \lambda_a} \rangle
\end{equation}
is related to \textit{normally hyperbolic invariant manyfold} (NHIM)
with $p$ unstable directions~\cite{Wiggins2001}. These are manifolds
which are unvariant under the dynamics and whose normal directions
have the structures of saddles, with exactly $p$ unstable
directions. To see this, consider an overdamped Langevin dynamics with
very small noise. Computing $Z_1$ amounts to selecting trajectories
that emerge from a saddle point with exactly one unstable direction
(see figure \ref{fig:flow}), the changes in distribution due to the
acceleration or deceleration along the trajectory is exactly
compensated by the factor $\Lambda_1$. All degrees of freedom but one
lie at the bottom of energy wells and $Z_1$ has stabilized the
\textit{unstable manifold} emanating from a NHIM with one unstable
direction. Similarly, computing $Z_2$, where we bias with
$\lambda_1+\lambda_2$, exactly compensates the area variations
incurred when leaving from a saddle with \textit{two} unstable
directions, all other degrees of freedom again lying at the bottom of
energy wells.  In general, computing $Z_p$, the bias with $\Lambda_p$,
stabilizes the $p$-dimensional unstable manifold emerging from
critical points with $p$ unstable directions.  What we have described
is the basis of the relation between supersymmetry and Morse theory
(\cite{witten,TTK}), expressed in the context of stochastic
equations. For dynamics more general than purely dissipative, the
structures stabilized by $Z_p$ are not restricted to unstable manifold
emerging from critical points, as we shall see below for Hamiltonian
dynamics.}

Another area where fluctuations of Lyapunov exponents are expected to
play a crucial role is that of intermittent dynamics. In such cases,
the coexistence in phase space of trajectories with different
chaoticities can be seen as the signature of a first order phase
transition. Measuring the free energy $\mu(\alpha,t)$ and order parameter
$\lambda(\alpha)$ in intermittent system is thus an important challenge.
To understand why intermittency can be related to first order
dynamical transitions, consider the following example. Suppose there
is a metastable structure, with a lifetime ${\cal{T}}$. Within this
structure, the largest Lyapunov exponent $\lambda$ takes an average
value $\lambda_{ms}$. On the other hand, the average of $\lambda$ over
the typical trajectories (the stable attractor) is, say,
$\lambda_{s}$.  We wish to calculate the large deviation function
$\langle e^{\alpha t \lambda} \rangle$, and to estimate which
trajectories contribute (see \cite{six}).  If {$\alpha
(\lambda_{ms}-\lambda_s)>0$}, the trajectories that belong to the
metastable structure will be favored by the bias $\alpha$ in the
measure.  The quantities $\alpha \lambda_{ms}$ and $\alpha \lambda_s$
are the biases per unit time attributed to trajectories in metastable
and stable basins, they have dimensions time$^{-1}$. This bias is
compensated by the time of escape from the metastable state ${\cal
  T}$, so that the condition for the trajectories in the metastable
state to dominate the large deviation computation is:
  \begin{equation} 
{\cal{T}} \alpha (\lambda_{ms} - \lambda_s) {\gg} 1
\end{equation}
 If $(\lambda_{ms} - \lambda_s)$ is extensive, and the metastable
 state is rather stable (${\cal{T}}$ is large), the transition takes
 place close to $\alpha=0$ \cite{FVW}.  It will be sharp, and first
 order.

\vspace{.25cm}

{\bf Outline}

\vspace{.25cm}

In what follows we shall present several applications of the
\textit{Lyapunov Weighted Dynamics} (LWD), a cloning algorithm that
allows to simulate a population of trajectories (or clones) with a
biased measure $P(\lambda,\tau) e^{\alpha\lambda
  \tau}$~\cite{Tailleur2007}. We will briefly review the numerical
method in section~\ref{sec:LWDIntro} while the technical details, the
underlying formalism and its extension to the case of several Lyapunov
exponents are presented last, in section~\ref{sec:method}, to make the
article more readable.

Then, we will show how the LWD can localize tiny regular island in
almost completely chaotic systems (section~\ref{sec:chaos}) and detect
the disappearance of the last regular islands steming from the
Lagrange points in the restricted gravitational three-body problem
(section~\ref{sec:Lagrange}). We will then show in section~\ref{separ}
that regular Hamiltonian systems develop positive Lyapunov exponents
when perturbed by a weak, additive noise and that trajectories with
atypical Lyapunov exponents are localized on interesting objects such
as separatrix, degenerate tori or unstable manifolds. We will then
turn to systems with large number of degrees of freedom, showing that
trajectories with atypically large $k$th Lyapunov exponent in the
Fermi-Pasta-Ulam chain typically involve $k$ chaotic breathers (see
section~\ref{sec:FPU}). Last, we will show in section~\ref{sec:LDFCTM}
that the LWD allows one to compute the dynamical free energy
$\mu(\alpha)$ and order parameter $\lambda(\alpha)$ in one-dimensional
coupled-map lattices.

\section{Lyapunov Weigthed Dynamics - presentation of the algorithm}
\label{sec:LWDIntro}
{For sake of clarity, let us consider a given one-dimensional
\textit{stochastic} dynamical system
\begin{equation}
  \label{eqn:dyn}
  \dot x = f[x(t)]
\end{equation}
where $f(x)$ is a complicated function that contains some noise
terms. Let us call $\eps$ a parameter controlling the noise intensity
so that $\eps=0$ is the deterministic limit of the system. To compute
Lyapunov exponents, one also need to consider the linearized dynamics
\begin{equation}
  \label{eqn:lindyn}
  \dot u = f'[x(t)] u
\end{equation}
From the solution of~\eqref{eqn:lindyn}, one can then compute the largest Lyapunov exponent 
\begin{equation}
  \lambda_1(t)=\frac 1 t \log\frac{|u(t)|}{|u(0)|}
\end{equation}

The LWD is a population Monte Carlo algorithm whose purpose is to
sample trajectories according to a modified measure so that a
trajectory $x(t)$ with a finite-time Lyapunov exponent $\lambda_1(t)$
is sampled with a probability
\begin{equation}
\label{eqn:tiltedmeasure}
  P_\alpha[x(t)]=\frac 1 {Z(\alpha,t)} P[x(t)] e^{\alpha t \lambda_1(t)}\qquad\mbox{with}\qquad Z(\alpha,t)=\int {\cal D}[x(t)]  P[x(t)] e^{\alpha t \lambda_1(t)}
\end{equation}
where $P[x(t)]$ is the probability to observe the trajectory $x(t)$
with the unbiased dynamics~\eqref{eqn:dyn}. The algorithm was
introduced in~\cite{Tailleur2007} and briefly reviewed
in~\cite{Giardina2011} (see also~\cite{Balint} where it was used to
localize atypical trajectories in convex billiards). It is presented in
full details in section~\ref{sec:LWD_Algo} after a quite extensive
discussion of the underlying formalism at the beginning of
section~\ref{sec:method}. Nevertheless, before we present several
applications of the algorithm in the sections \ref{sec:chaos} to
\ref{sec:spatext}, we briefly present the method below and try to give
hints about why it is efficient and where are its pitfalls.

The principle of the algorithm is as follows. One simulates $N_c$
copies $(x_i(t),u_i(t))$ of the system, also called ``clones'', with
the unbiased dynamics~\eqref{eqn:dyn} and~\eqref{eqn:lindyn}. Then,
at every time step, one computes for each clone $i$
\begin{equation}
  s_i(t)=\frac{|u_i(t)|}{|u_i(t-\dt)|}
\end{equation}
The clone $i$ is then replaced, on average, by $s_i(t)^\alpha$ copies. If the
dynamical system were deterministic ($\eps=0$), all the $s_i(t)^\alpha$
copies would then follow the same trajectory and a given clone would
yield, after a time $t$, a total number of copies given by:
\begin{equation}
  \prod_{n=1}^{t/\dt} s_i(n \dt)^\alpha = \prod_{n=1}^{t/\dt}
  \frac{|u_i(n\dt)|^\alpha}{|u_i((n-1)\dt)|^\alpha} = \frac{|u_i(t)|^\alpha}{|u_i(0)|^\alpha}=
  e^{\alpha \lambda_1 t}
\end{equation}
If one were to pick a clone at random in the population at time $t$,
the probability to choose a trajectory $x(t)$ would thus be
proportional to $e^{\alpha \lambda_1 t} P[x(t)]$, where $P[x(t)]$ is
the probability that the dynamics \eqref{eqn:dyn} produces the
trajectory $x(t)$. This thus results in the biased
measure~\eqref{eqn:tiltedmeasure}. Of course, simulating many times
the same trajectory is not enhancing the quality of the sampling and
will not reveal any rare trajectories that brute-force sampling would
miss. This is where the stochasticity of $f[x(t)]$ comes into play:
when $\eps\neq 0$, all copies of a given clone made at a time $t'$
will follow different trajectories for $t>t'$. Since the cloning is
larger for trajectories that carry a large weight, their vicinity is
well sampled by the algorithm. Conversely, trajectories with small
weights rapidly die out and no computing power is spent on their
simulation. In practice, there are some differences---mostly due to
numerical efficiency---between the LWD and the algorithm presented
above (for instance, the number of clones $N_c$ is kept constant in
LWD), which are detailed in section~\ref{sec:LWD_Algo}.

In various parts of this paper, noise is not part of the original
problem, and it is added only to let the different clones diffuse
independently, to enhance the sampling of the algorithm.  In those
cases, the level of noise $\eps$ has to be taken as small as possible
and one should ensure that the $\eps\to 0$ limit is not
singular. Finally, $\eps$ can be taken to zero at the end of the run
to check that the atypical trajectories that have been discovered are
indeed solution of the equations of motion of the original system.

When the system conserves energy, it may be desired to search for
trajectories within a specified energy shell -- and the same can be
said of other constants of motion, if present. It is then useful to
consider energy-conserving noise. This may be implemented in a number
of ways, such as rotating the velocity vector randomly, or, more
generally proposing a small random change of the coordinates and
projecting them back onto the energy shell (see~\ref{sec:bruitcons} for a
momentum- and energy-conserving noise in Hamiltonian systems).

The values of $\alpha$ and $\eps$ play roles analogous to the inverse
temperature and the step size in a standard Metropolis Monte Carlo
programme. If $\alpha$ is too large, the system may stay trapped in a
locally favoured but globally non-optimal situation. If $\alpha$ is
too small, the noise might make the system miss a convenient but small
structure. The same is true about the noise. Too small a noise makes
the programme run very slowly, as the clones do not have time to
evolve differently, and the selection process is inefficient. Too
large a noise makes the system miss structures that are dynamically
unstable: this is quite crucial when one is dealing with highly
chaotic systems.

Last, one of the most difficult parameter to choose is the number of
clones used in the simulations~\cite{Hurtado}. Typically, small numbers ($N_c \sim
100$) will suffice to locate atypical trajectories (say an integrable
island in a chaotic sea) whereas much larger populations may be needed
if one wants to compute accurately the averages $\mu(\alpha,t)$ or
$\lambda_1(\alpha,t)$ (see section~\ref{sec:LDFCTM} for a discussion
of the convergence of the algorithm as $N_c\to \infty$).}

\section{Testing the chaoticity of a system}
\label{sec:chaos}

An interesting question in dynamical systems concerns the existence of
\textit{smooth} potentials having completely chaotic, fully ergodic
dynamics.  A candidate for this was for many years the
Hamiltonian~\cite{candidate}
\begin{equation}
{\H = \frac 12 (p_x^2 + p_y^2 + x^2 y^2)}
\end{equation}
The expectation actually turned out not to be justified, as was proven
by Dahlqvist \& Russberg~\cite{dahlqvist1990}, who managed, using an
elaborate strategy, to find a tiny (area $\sim 10^{-7}$) regular
island.

This seems like a good benchmark case to test the Lyapunov Weighted
Dynamics~\cite{predrag}.  In order to select low-Lyapunov
trajectories, we use LWD to bias the measure on trajectories by 
$  e^{\alpha t \lambda(t)} $ with $\alpha < 0$ and $\lambda$ the largest Lyapunov exponent.

The energy may be taken by rescaling to be $E=\frac 12$.  A small
amount of noise is added to the dynamics, which rotates the
$(p_x,p_y)$ vector by a random angle with typical value $\sim
\sqrt{2\eps \delta t}$, where $\delta t$ is the integration step, and
$\eps$ {controls the amplitude of this energy-conserving
  noise. As the simulation proceeds, it is slowly taken to zero, much
  in the same way as in standard simulated annealing}.  Starting from
a random point, the program falls in a few {million
  steps} in the island found by Dahlqvist et Russberg, which shows
that it is a relatively {large target for the programme}.
We next try to find a new island, but for this we have to avoid
falling there.  This may be easily done by killing all walkers that
enter a region including this attractor (for example, the rectangle
$3.13 \le x\le 3.16$ and $-0.006\le p_x \le 0.004$).

\begin{figure}[h!]
\begin{center}
\includegraphics[width=.6\textwidth]{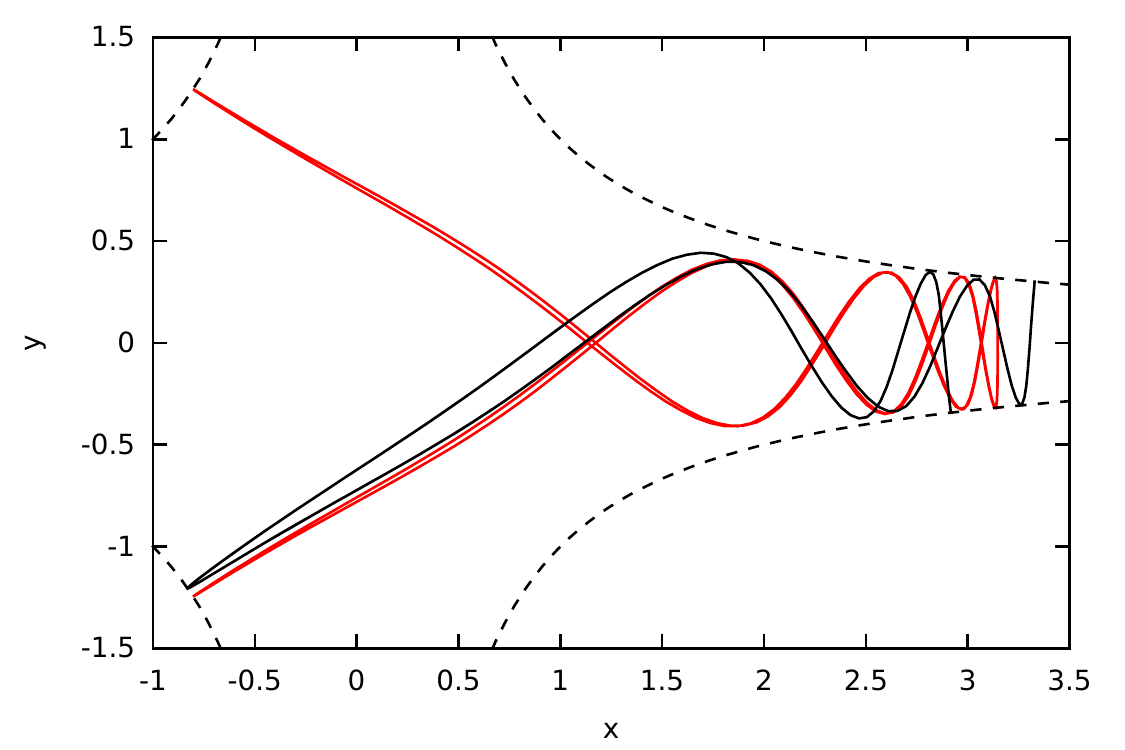}
\end{center}
\caption{{A periodic orbit found by the clones (black curve). The
    integrable island also contains the image of this orbit by the
    transformation $y\to-y$ as well as the rotations of these orbits
    by $\pi/2$, $\pi$ and $3\pi/2$. The trajectory retraces its steps
    at its ends at the level indicated with dashed lines, where its
    velocity goes to zero. Dashed curves $x^2 y^2=1$ enclose the
    energetically allowed area. The red curve is the shorter periodic
    orbit found by Dahlqvist \& Russberg~\cite{dahlqvist1990}.}}
\label{fig:x2y2orbit}
\end{figure}

\begin{figure}[h!]
\begin{center}
\includegraphics[width=.6\textwidth]{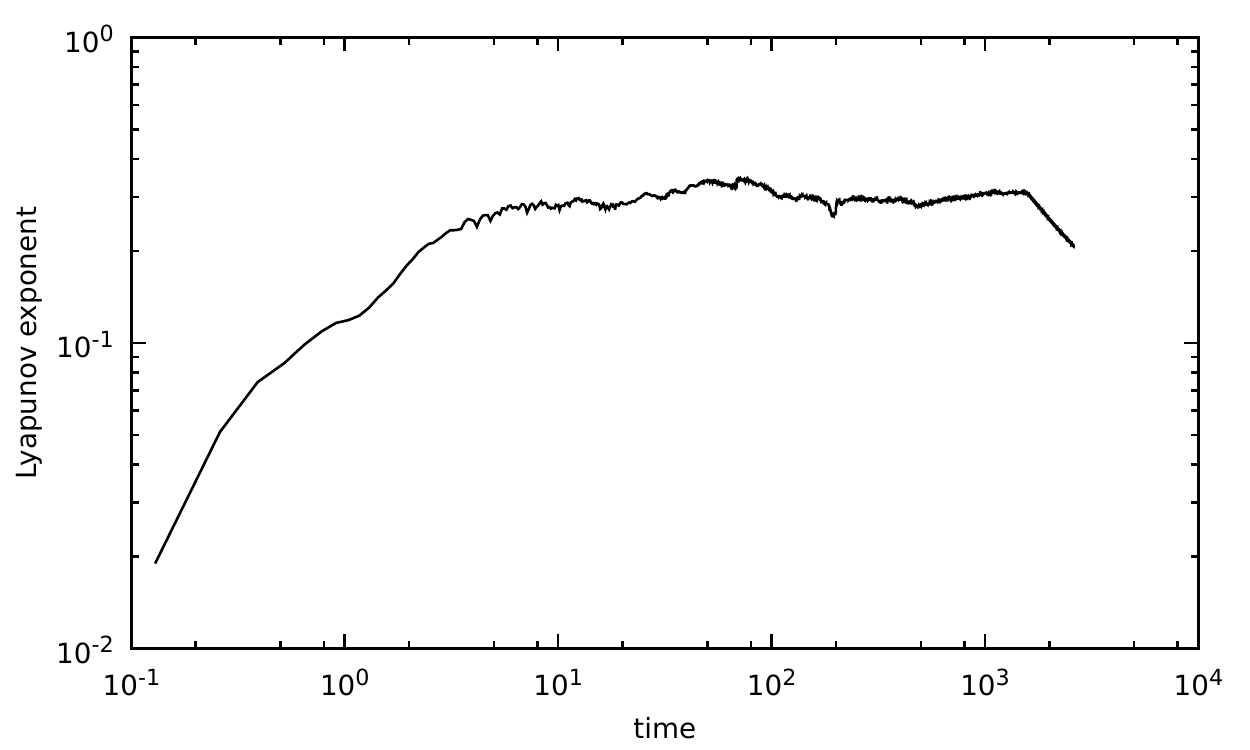}
\end{center}
\caption{{A logarithmic plot of the finite-time Lyapunov exponent up to time $t$.
The tell-tale sign that a region with negligible Lyapunov exponent has been found is the beginning
of a $1/t$ regime: here it happens around $t=1500$.}
 We used $1000$ clones, $\alpha=-1$, $\eps=10^{-10}$.  }\label{fig:x2y2monitoring}
\end{figure}

When this is done, the clones find a new island, \textit{with area of
  the order of $10^{-10}$}. A trajectory within this island is
depicted in figure \ref{fig:x2y2orbit}.  One knows that an island has
been found by monitoring the finite time Lyapunov exponent (cf. figure
\ref{fig:x2y2monitoring}): the beginning of the familiar $\frac{1}{t}$
regime is a signal that the Lyapunov is approaching
zero~\cite{Tailleur2007}. Note that this strategy works in all
dimensionalities, and that it requires no special knowledge of the
system. {Let us note that the noise amplitude (here
  $\eps=10^{-10}$) has to be small enough that the clones do not
  ``step over'' the integrable island in one time-step. The smaller
  the island one wants to detect, the smaller $\eps$ should
  be---whence the aforementioned annealing in $\eps$ done in the
  simulations.}

\if{Inspired by this exercise, it seems that one could find further
families, by adding an extra `bounce' on the two ends of the
trajectory.}\fi

\section{Stability of the  Lagrange points L4 and L5}
\label{sec:Lagrange}
{Consider the restricted gravitational three-body problem
  {(see e.g. \cite{3bp})}, with two rescaled masses $1-\mu$ and
  $\mu$ (e.g. Sun and Earth), and a small third body of negligible
  mass {(such as a satellite or asteroid).}

When the large bodies are moving in a circular orbit, there are points
where the small mass may be placed such that in a rotating frame the
three bodies are seen as stationary. These are the so-called Lagrange
points L1-L5, of which L4 and L5 are stable. When the trajectories of
the large bodies are elliptic, with an excentricity $e>0$, they are
not anymore stationary in this rotating frame, but Lagrange points
still exist. For instance, when $\mu$ and $e$ are sufficiently small, it
turns out that the vertex of the (time-dependent) equilateral triangle
defined by the three bodies is a stable trajectory where the small
mass moves in phase with the large masses. Many works starting from
the one of Danby~\cite{danby1964} where devoted to studying this
situation.

Denoting the positions of the larger bodies in the rotating frame
${\bol r}_1(t)$ and ${\bol r}_2(t)$, and normalizing lengths and
times so that the period of a circular orbit is $2\pi$, the
Hamiltonian \textit{for the smallest particle} in the rotating frame
is:
\begin{equation}
{\H(\bol r, \bol p) = \frac12 (\bol p - \bol A)^2 + V(\bol r)}
\label{jacobi}
\end{equation}
with $\bol A = (-y, x, 0)$ providing the Coriolis force and $V =
-\frac 1 2 r^2 - (1-\mu)/|\bol r - \bol r_1| - \mu/|\bol r - \bol r_2|$. 

When the orbits of the large bodies are circular, $\bol r_1(t)=\bol
r_1^\mu\equiv(-\mu,0,0)$ and $\bol r_2(t)=\bol {r_2^\mu}\equiv
(1-\mu,0,0)$. When $e>0$, $\bol r_1(t)$ and $\bol r_2(t)$ are
obtained by solving the Kepler problem; the larger masses librate around $\bol r_1^\mu$
and $\bol r_2^\mu$ with an amplitude that is larger for larger
excentricities. In both cases, the Hamiltonian (\ref{jacobi})
is the only constant of motion. This means that the restricted
two-degree of freedom problem, where the small particle moves in the
same plane as the large ones, is not integrable.  It is, in fact, a
mixed system with both regular and chaotic regions, the former
containing the Lagrange points L4 and L5. When the eccentricity $e$ increases, or when the masses become
comparable, the system is further perturbed by the apparent motion in
the rotating frame of the large masses. The effect is to reduce
further the regular islands. In particular, the regular regions
containing L4 and L5 shrink and develop resonances, until they
eventually disappear.
 
We have used the LWD to find the part of the $(\mu,e)$ plane where
regular regions still exist around $L_4$ and $L_5$. We have used a
population of 100 clones and applied a strong bias {
  ($\alpha = -10^{4}$)}, favoring stable orbits. We used such a strong
bias because as soon as the clones leave the regular vicinity of the
Lagrange points, they quickly diverge to infinity and have a zero
probabilty of coming back. A strong bias thus ensures that some clones
will always remain within the integrable region. {Just as
  in the previous section, the noise required to use LWD is provided
  by rotating the velocity $ (\bol p - \bol A)$ by a weak random angle
  at each time step, thus conserving the Jacobi constant
  (\ref{jacobi}).  The typical value of this angle is weak: $\eps = 10^{-15}$,
  and it allows the different clones to search for the best
  situation.}  {For each value of $\mu$, we start all the clones
    at the Lagrange point L4 and adiabatically increase the
    eccentricity ($e(t) = \dot e t$ with $\dot e=10^{-5} \ll 1$),
    until none of the clones is able to remain in the integrable
    islands and they all diffuse to infinity. This yields the first
    value $e_c$ at which the islands around L4 disappear. We have
    checked that the results do not depend on the parameter $\dot e$,
    as long as it is much smaller than all typical time derivatives in
    the natural dynamics (which are of order one).
Note that such a protocol does not give access to the re-entrant part of
the stability region (rightmost gray part of the figure).
In order to obtain the two isolated points on the figure, we slowly
increased not only $e$ but $\mu$ as well so that these points
correspond to the first time the clones leave the integrable island.}
In this way we have obtained automatically (and blindly) the stability
region of figure \ref{fig:L4_stability_diagram}. Using a linear
stability analysis of $L_4$ and $L_5$, Danby~\cite{danby1964} was able
to predict a lower bound of this stability region, which is very close
to the one we observe numerically. This suggests that the regular
region surrounding $L_4$ and $L_5$ quickly disappears when they become
linearly unstable.

\begin{figure}[h!]
\begin{center}
\includegraphics{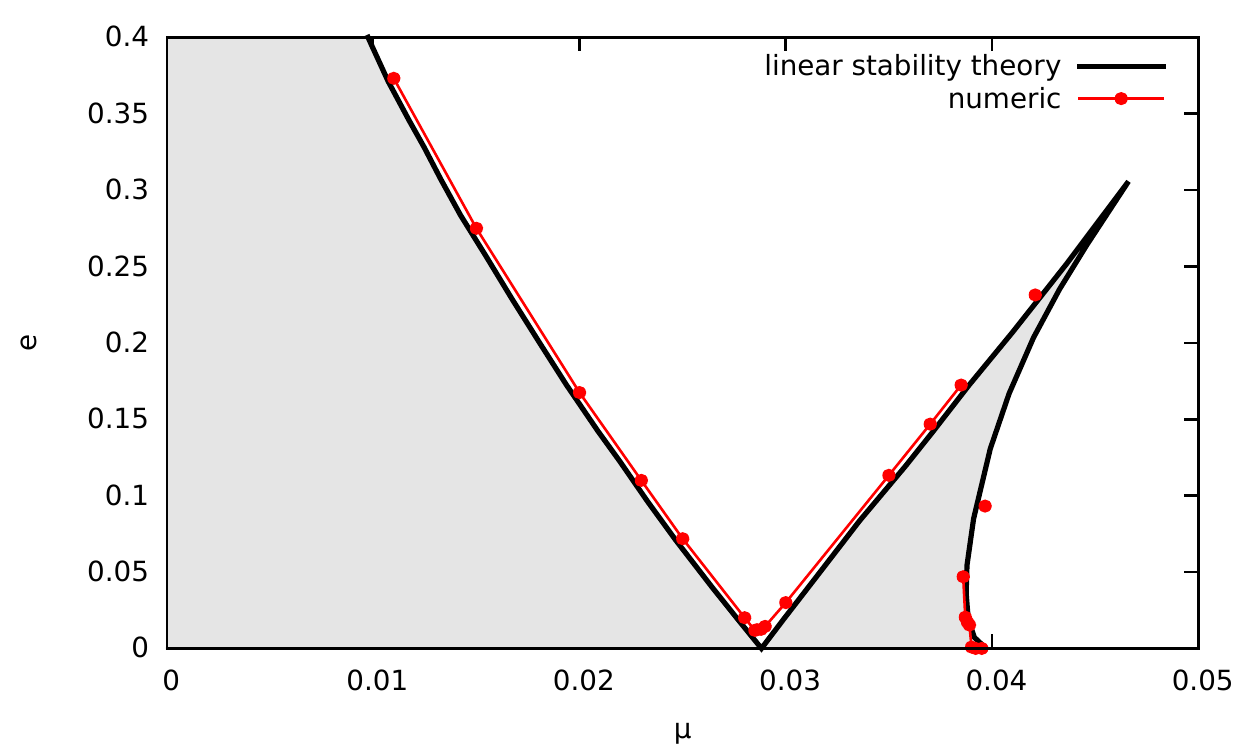}
\end{center}
\caption{{Stability region (in gray) of the Lagrange points, in
    terms of eccentricity and mass ratio. Boundaries in red are
    obtained by the LWD while the solid lines are the predictions of
    Danby~\cite{danby1964}, based on a linear stability analysis of
    $L_4$ and $L_5$. The red points connected by a solid line were
    obtained by increasing the eccentricity from $e=0$ until none of
    the clones is able to remain in the integrable island. The two
    disconnected points were obtained by slowly increasing both $e$
    and $\mu$ so that the first time the clones leave the stability
    region is in the re-entrant region. The number of clones is
    $N_c=100$, the bias is $\alpha=-10^4$ and the level of noise is
    set by $\eps=10^{-15}$.}}\label{fig:L4_stability_diagram}
\end{figure}}

\section{The case of integrable  dynamical systems}
\label{separ}
\subsection{Stochastic perturbation of integrable systems}
\label{NILE}
An integrable system has obviously all its Lyapunov exponents equal to
zero.  However, a somewhat surprising fact is that if we perturb such
a system with a weak, additive white noise, it acquires a non-zero
Lyapunov exponent, defined as the rate of separation of nearby
trajectories \textit{subjected to the same noise realization}.

Let us consider first a  problem with one degree of freedom. For example, for the Hamiltonian 
\begin{equation}
{\H= \frac{p^2}{2}+ V(q)}
\end{equation}
the equations of motion, once the perturbation $\eta(t)$ is added,  read:
 \begin{eqnarray}
\dot q &=& p \nonumber \\
\dot p &=& -\frac{dV}{dq} + \eta(t)
\end{eqnarray}
Here, $\eta(t)$ is a white noise with variance $\langle \eta(t)
\eta(t') \rangle = 2 \eps \delta(t-t')$.  For small $\eps$, energy drifts
slowly, and one may prove that there is exponential separation of
trajectories \textit{along the line of constant energy}, induced by
the noise \cite{KDK}.  The origin of this surprising phenomenon is
that noise allows some trajectories to move into a faster moving orbit
of slightly higher energy, and then come back, {a
  phenomenon that is related to, but distinct from, Taylor's diffusion
  in hydrodynamics \cite{Taylor}.} {It is, on the other
  hand, quite different from the usual form of `Noise Induced Chaos',
  which is originated by the noise-induced occasional visits a system
  makes of unstable chaotic regions \cite{tel}, there being in our
  case no chaotic regions at all.}

{Clearly, for a harmonic oscillator the present mechanism cannot
  work, as all orbits have the same period, regardless of the
  energy. As we will show elsewhere~\cite{KDK}, when the variance of
  the noise $\eps\to 0$, the diffusion occurs mostly tangentially to
  the tori on time-scales up to $t \sim \eps^{-1/3}$. The induced
  divergence of two nearby trajectories is exponential and the
  corresponding Lyapunov exponent can be computed
  explicitly~\cite{KDK}:}
\begin{equation}
{\lambda =  \frac{\sqrt\pi}{\Gamma(\frac16)} \left(
\frac 3 2 \eps
  \overline{\left(\frac{\partial^2 q(I,\phi)}{\partial\phi^2}\right)^2}
 \left( {\frac{\d^2 \H(I)}{\d I^2}}\right)^2
\right)^{1/3}
= \left(\frac32\right)^{1/3} \frac{\sqrt\pi}{\Gamma(\frac16)}
\left(
  \eps
  \overline{(\ddot q)^2}
  \left(\frac1\omega \frac{d\omega}{d E}\right)^2
\right)^{1/3}}
\label{formu}
\end{equation}
where $I$ and $\phi$ are the action and angle variables of the
problem, {$\omega= \dot \phi=\frac{\partial \H}{\partial I}$} is the
frequency.  The variable $q(t)$ is, in the absence of noise, a periodic function of $\phi$,
  the overbar in (\ref{formu}) denotes average over a
cycle {of the noiseless dynamics}.  {The Lyapunov time is proportional to $\eps^{-\frac 13}$, while the diffusion away from a torus takes times of order
$\eps^{-1}$: for small $\eps$ there is then a wide range of times $\eps^{-\frac 13 } \ll t \ll \eps^{-1}$ where the definition of a Lyapunov
exponent associated with a torus makes sense.}

{An interesting limit of equation (\ref{formu}) corresponds to the
approach of a separatrix of energy $E_s$, where the frequency $\omega$
vanishes.}  One may easily check that, quite generally, in terms
of $\Delta E = E-E_s$:
\begin{equation}
\omega(\Delta E \to 0) \approx -\frac\pi{\log\Delta E} \to 0
\end{equation}
so that:
\begin{equation}
{\frac{d^2 \H}{dI^2} = \omega \frac{d\omega}{dE} \propto \frac1{\Delta E |\log\Delta E|^3} \to \infty}
\end{equation}
Hence, as one may have expected, the Lyapunov exponent becomes large
as one approaches a separatrix, on which it becomes larger than {${\cal
  O}(\eps^{1/3})$}.

\begin{figure}[h!]
  \centering
  \includegraphics{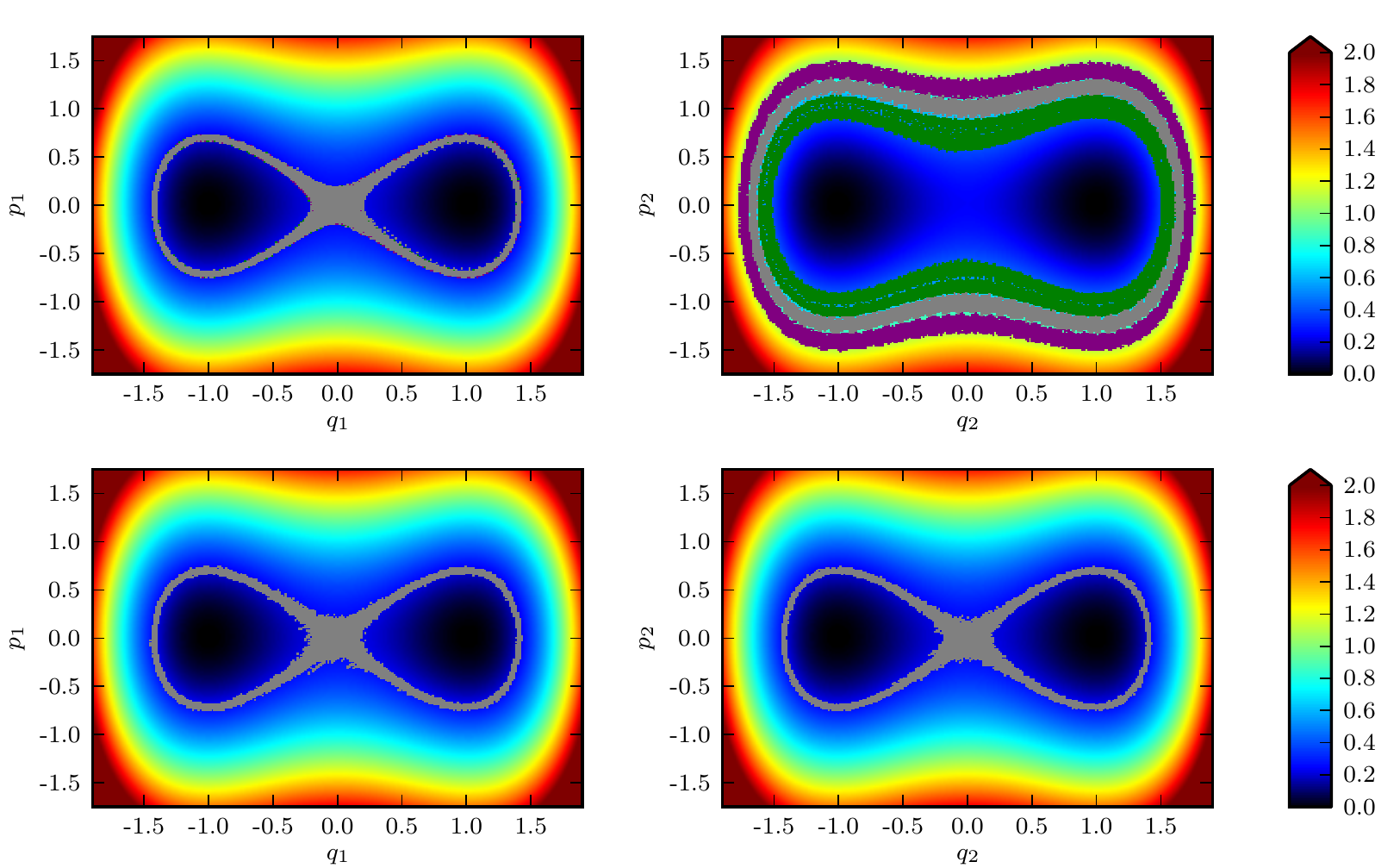}
  
  \caption{LWD with $2\,000$ clones for a system defined
    by~\eqref{eqn:dyn37} and $\eps=10^{-5}$. {\bf Top}: $\alpha_1=1$
    and $\alpha_{i\geq 2}=0$. Position of the clones for $t\sim 5\,
    000$ (dark green), $t\sim 10\, 000$ (purple) and $t\sim 15\,000$
    (gray). {\bf Bottom}: $\alpha_{1,2}=1$ and
    $\alpha_{3,4}=0$. Stationary positions of the clones (here $t\sim
    2\,000$ in gray).}\label{fig:separatice}
\end{figure}

\subsection{The special case $\alpha=1$}
{In order to see what kind of structures dominates the dynamical
partition function $Z_p=\langle
e^{(\lambda_1+\dots+\lambda_p)t}\rangle$, let us consider a system of
two degrees of freedom that consists of a cartesian product of two
systems of one degree of freedom. {In figure~\ref{fig:separatice}
  we show the result of LWD for a cartesian product of two double
  wells $\H(\bol q,\bol p)=\sum_{i=1,2}p_i^2/2+(q_i^2-1)^2/4$, with
  additive Gaussian white noise of variance $2\eps$:
\begin{equation}\label{eqn:dyn37}
  \dot q_i = p_i\qquad \dot p_i=-q_i (q_i^2-1)+\sqrt{2 \eps}\, \eta_i
\end{equation}
This system has two \textit{normally hyperbolic invariant manifold}
(NHIM) with one unstable direction defined by $q_1=p_1=0$ for the
first one and $q_2=p_2=0$ for the second one. They correspond to
the cartesian products between the flat measure~\footnote{Note that
  \eqref{eqn:dyn37} is the limit of a Kramers equation with friction
  $\gamma p$ and temperature $kT$ when $\gamma\to 0$ and
  $kT\to\infty$ with $\eps=\gamma kT$ constant. Without cloning, the
  steady-state measure is thus the infinite temperature limit of a
  Boltzmann weight, i.e. the flat measure.} over one double well and
the saddle point of the other double well. It also has one NHIM with
two unstable directions defined by $q_1=q_2=p_1=p_2=0$ which is the
cartesian product of the two saddle-points. The walkers are biased to
search for trajectories with either an atypically large $\lambda_1$
($\alpha_1=1$ and $\alpha_{i\geq 2}=0$) or an atypically large sum
$\lambda_1+\lambda_2$ ($\alpha_{1}=\alpha_{2}=1$,
$\alpha_3=\alpha_4=0$).}

When we bias the walkers with $e^{\lambda_1 t}$, we stabilize the
unstable manifold of one NHIM with one unstable direction: the clones
are localized on the separatrix in one of the double wells while they
diffuse freely in the other one (since the noise does not conserve the
total energy). On the other hand, $e^{(\lambda_1+\lambda_2) t}$
stabilizes the unstable manifold of the NHIM with two unstable
directions, i.e. the cartesian product of the two separatrices.

\subsection{The most regular trajectories ($\alpha<0$)}
{As already mentioned above, choosing $\alpha=1$ is such that the
walkers populate the separatrix \textit{uniformly}: the stretching and
compressing in times of acceleration and deceleration are exactly
compensated by the cloning rate~\cite{TTK}. In the following we
consider the converse case with $\alpha<0$.

In a Hamiltonian system, the motion in a regular region is constrained
to a torus, labeled by the constants of motion $I_a$, and spanned by
the dynamic curves $\dot \theta_a= \frac{\partial \H}{\partial I_a}$.
By changing the values of the $I_a$, one moves to different nested
tori.  Consider for definiteness a system with two degrees of freedom
with two constants of motion $I_1,\,I_2$ and the corresponding angles
$\theta_1,\,\theta_2$ which span the two-dimensional surface of the
torus. The innermost of these tori reduces to a ring, spanned by only
one of the two angles, say $\theta_1$. In the vicinity of this ring,
the dynamics of $I_2, \theta_2$ generically reduces to a harmonic
oscillator. In mixed systems, such a degenerate trajectory corresponds
to the center of an integrable island.

Let us now show how the LWD can be used to find such a degenerate
structure. As shown in section~\ref{NILE}, the noise-induced Lyapunov
exponent in integrable systems come from the anharmonicities that make
$\H''(I)\neq 0$. This may be seen in equation (\ref{formu}) by
checking that in the case of a harmonic oscillator, for which $\H=I$,
the Lyapunov exponent vanishes. \if{Now, if one expands $V(q)$ around
  a local minimum, generically the first order is quadratic, so that
  the orbits with smaller energy are also those with smaller Lyapunov
  exponent.}\fi Looking for the trajectory with the smallest Lyapunov
exponent in the presence of noise, amounts in fact to looking for the
``most harmonic'' of trajectories. In the case of nested tori, this
generically corresponds to the innermost trajectory, which is
degenerate.

As an example, consider the restricted three-body problem {with
  $\mu=0.1$} and the large masses performing circular orbits.  As
mentioned above, the system has many regular islands, that correspond
to non-chaotic orbits, many of which have been classified
\cite{barrabes}. Figure \ref{fig:degeneratetori} concerns one such
regular island: the typical trajectories in there, such as the red
one, are rather complicated in real space.  Now consider the
degenerate ``innermost'' torus in the island, in black: it corresponds
to the ``flower'' structure which is stationary in the rotating
frame. Indeed, we recognize every other regular trajectories of the
island as essentially this trajectory, with, in addition, librations
of this basic pattern.  Clearly, even for analytic purposes, it seems
useful to be able to locate such a structure. (Their role is, by the
way, reminiscent of the role played by ``inherent structures'' in glassy
problems {where all the vibrational motion is eliminated
  by treating as single configuration all points belonging to a same
  basin of attraction of the zero-temperature dynamics}). Furthermore,
while it is possible to ``guess'' that the red trajectory of
figure~\ref{fig:degeneratetori} corresponds to librations around the
black one, this would be impossible in higher dimensions while the LWD
would nevertheless have no trouble finding the proper degenerate
trajectory. Figure~\ref{fig:driftingclones} shows, for the example
above, the clones migrating to the center of an integrable island, when
selected for low Lyapunov exponent.}

\begin{figure}[h!]
\begin{center}
\includegraphics{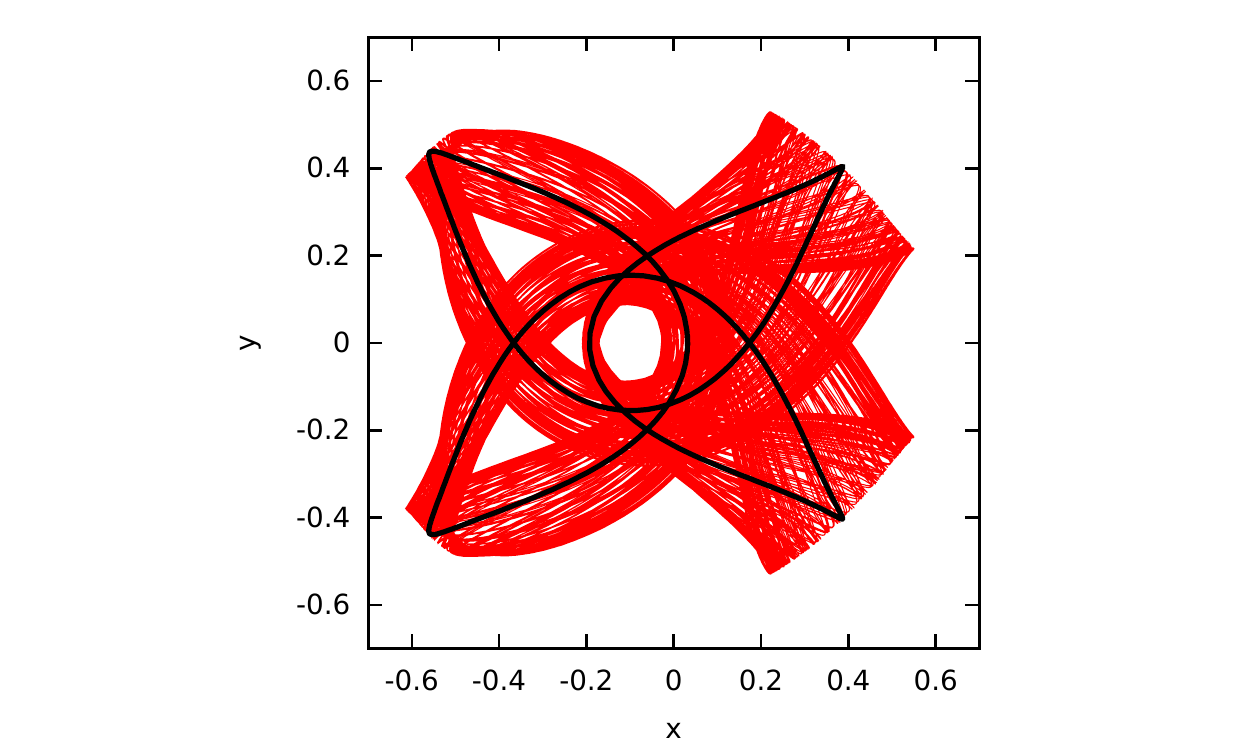}
\end{center}
\caption{Simulation of the restricted three-body problem~\eqref{jacobi}
  with an eccentricity $e=0$ and $\mu=0.1$. Red: quasi-periodic stable
  orbit.  Black: the periodic orbit at the center of the corresponding
  island of stability.  }\label{fig:degeneratetori}
\end{figure}

{Higher dimensional structures may be selected for an integrable problem
with more degrees of freedom. For instance, minimizing the sum of the
smallest $p$ positive exponents $\lambda_1+...+\lambda_p$ leads to
tori that are $p$ times degenerate (i.e. they are $N-p$ dimensional).
These considerations might be interesting to apply to extended
integrable systems, such as the Toda lattice, because it might give a
method to construct soliton solutions.}
\begin{figure}[h!]
\begin{center}
\includegraphics{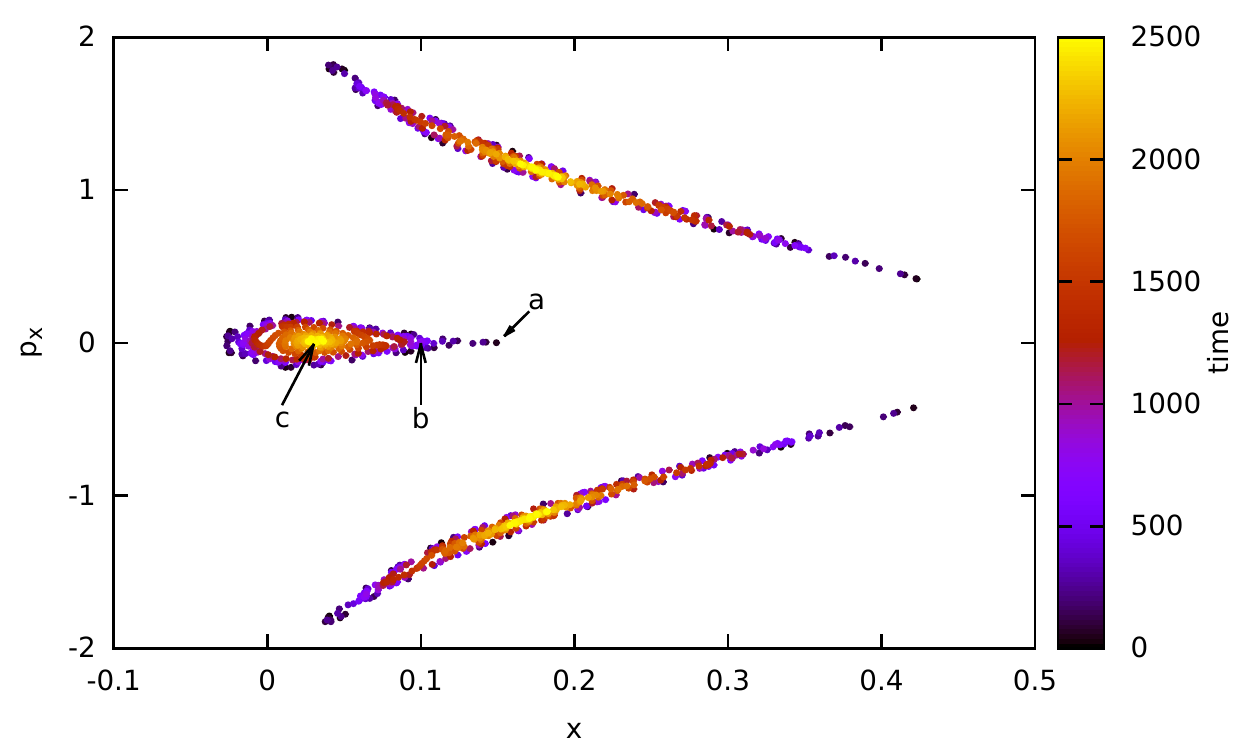}
\end{center}
\caption{{Poincar\'e section $(y=0,E=\H(L_2))$ corresponding to
    figure~\ref{fig:degeneratetori}. The clones are drifting from the
    initial position (a) to a nearby integrable island (b corresponds
    to the red trajectory of fig~\ref{fig:degeneratetori}).  They then
    migrate to the center of the island (c corresponds to the black
    trajectory of figure~\ref{fig:degeneratetori}).  Here $\H$ is
    given by~\eqref{jacobi}, we used 100 clones, $\alpha = -10^{6}$,
    and a noise intensity $\eps = 10^{-8}$.  The color code shows the
    time spent since the beginning of the simulation.}}
\label{fig:driftingclones}
\end{figure}

\section{Spatially extended systems}
\label{sec:spatext}
\subsection{Fluctuations of several Lyapunov exponents: the Fermi-Pasta-Ulam chain}
\label{sec:FPU}
\vspace{.5cm}

Let us now turn towards spatially extended systems and enter the
domain of condensed matter by considering the Fermi-Pasta-Ulam-Tsingou
chain defined by the Hamiltonian
\begin{equation}
\label{eqn:FPUHam}
  \H=\sum_{i=1}^L \Big[ \frac{p_i^2}2+ \frac {(x_{i+1}-x_i)^2}2 + \beta\frac {(x_{i+1}-x_i)^4}4\Big]
\end{equation}
with $x_{L+1}=x_1$. This corresponds to a chain of $L$ particles
coupled with anharmonic springs. In the $\beta=0$ limit, the system is
integrable and the chaoticity increases as $\beta$ and $\H$
increase~\cite{Dauxois1997}. In this paper, we consider the case
$\beta=0.1$ and $\H=L$. In previous
studies~\cite{Tailleur2007,Giardina2011} we showed that biasing this
system in favor of chaotic trajectories reveals chaotic breathers,
whereas regular trajectories correspond to a gas of solitons
propagating ballistically through the system. The unbiased case
corresponds to a mixture of short-lived solitons and breathers
superimposed with thermal fluctuations.

Two questions arise naturally when considering spatially extended
systems. First, what happens when we consider Lyapunov exponents
beyond the largest?  Second, how do the fluctuations scale with the
system size? Answering the latter in the FPU case is a difficult task
since even the average values already converge very slowly with the
system size~\cite{Searles1997}. Below, we thus address the former
question. {When using the LWD to sample the large deviations of
  the Hamiltonian dynamics associated with~\eqref{eqn:FPUHam}, we use
  the energy- and impulsion-conserving noise described
  in~\ref{sec:bruitcons}. Since the chaoticity of the FPU chain
  depends on the energy of the oscillators, it is indeed quite important to
  prevent the system from decreasing its chaoticity by either
  decreasing its total energy or by using all its energy to create a
  uniform rotation of all the oscillators.}

In~\cite{Ruffo1} it was shown that above a certain energy, a modulational
instability of the FPU chain leads to the formation of a chaotic
breather: the energy condensates on a small number of degrees of
freedom and the largest Lyapunov exponent increases
substantially. In~\cite{Tailleur2007} we showed that these chaotic
  breathers are the trajectories that contribute the most to
  $Z(\alpha,t)$ for large positive $\alpha$.

As we can see in figure~\ref{FPU1b}, the appearance of a chaotic
breather creates a gap in the Lyapunov spectrum, with the largest
Lyapunov exponent increasing by a factor 5, and the second one by a
factor 2. The plot of figure~\ref{FPU1b} includes cloning and noise,
so that one may think that the individual trajectories differ
completely from the underlying FPU dynamics, but this is not so: using
the positions and momenta of a given clone as an initial condition for
a numerical simulation without cloning and noise shows that the
breathers are indeed a proper solution of the FPU chain. As explained
in the introduction, the cloning simply stabilizes a metastable
solution of the equations of motion. {Since clones are constantly
  killed or copied as time goes one, one could be surprised that the
  trajectory in figure~\ref{FPU1b} does not show clear cloning events
  where, for instance, the breather would suddenly disappear and
  reappear few hundreds of site further. This can happen but is quite
  rare because, once a breather has been found, all the clones in a
  small population tend to be very similar. This ``degeneracy'' of the
  clone population, that sample the vicinity of \textit{one}
  interesting trajectory while many exist, is one of the reason why in
  section~\ref{sec:LDFCTM} we will need much larger population to
  accurately compute the average $Z(\alpha,t)=\langle e^{\alpha
    \lambda_1 t} \rangle$.}

The few first tangent vectors are localized on the breather, and one can
thus wonder what happens when we require that other Lyapunov exponents
be large. Either some internal degrees of freedom of the breather can
be excited, leading to an increase in Lyapunov exponents beyond
$\lambda_1$, or the system may divide the energy of the breather into
several breathers, each of them corresponding to one atypically large
Lyapunov exponent. A trajectory containing $k$ breathers is thus a natural
candidate for large deviations of the $k$ first Lyapunov exponents
but, as was shown in~\cite{Ruffo1}, the breathers have a tendency to
merge upon collision which makes multi-breathers solution unstable.

\begin{figure}
\includegraphics{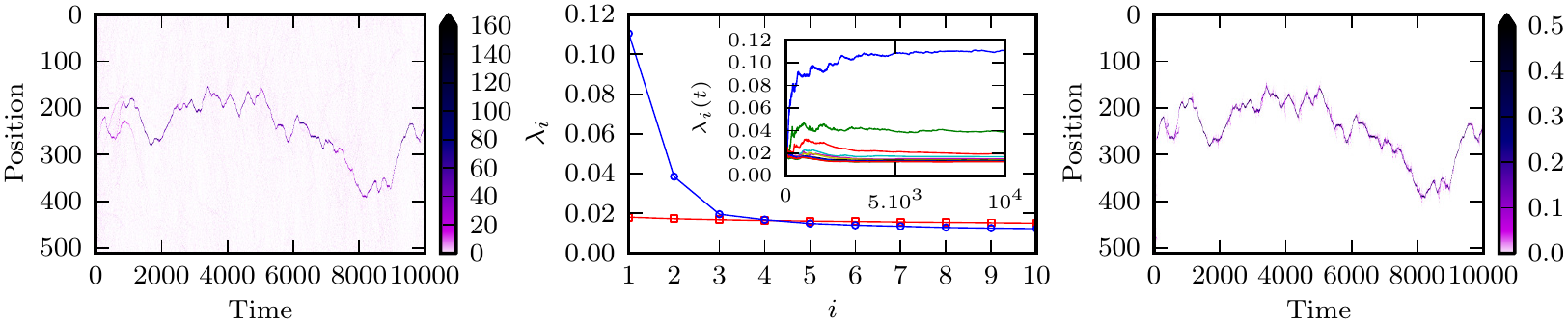}
  \caption{Simulation of a FPU chain with the Lyapunov Weighted
    Dynamics and parameters $L=512$, $\alpha=5L$, 200 clones,
    $dt=10^{-2}$, $\eps=10^{-3}$. {\bf Left}: Energy of each site (for
    one clone) as a function of time. {\bf Centre}: $\lambda_i$ vs $i$
    for $i\in [1,10]$ at $t=10^4$ for the biased (blue) and unbiased
    (red) case. The largest and second largest exponents are clearly
    larger in the presence of a breather. The inset shows
    $\lambda_i(t)$ for $t\leq 10^4$ for $i\in [1,10]$. {\bf Right}:
    $v_{q_i}^2+v_{p_i}^2$ as a function of time and lattice site $i$
    for the first tangent vector, which is shown to be localized along
    the breather.}
\label{FPU1b}
\end{figure}

As we show here, the breather-breather interaction is actually more
complicated and multi-breather solutions indeed exist. In
figure~\ref{FPU2b} we show that biasing with $\alpha_1=\alpha_{i\geq
  3}=0$ and $\alpha_2>0$ favors the appearance of two
breathers. Another Lyapunov exponent, $\lambda_2$, increases and the
gap is now between $\lambda_2$ and $\lambda_3$. The same is true for
$\lambda_3$ and $\lambda_4$ that generate solutions with 3 or 4
breathers (see figure~\ref{FPU2b}). {A large $\lambda_2$ also
  implies a large $\lambda_1$ since $\lambda_1\geq \lambda_2$ by
  construction. Using $\alpha_1>0 $ and $\alpha_2>0$ can thus also
  produce two breathers. However, as can be seen in
  figure~\ref{FPU1b}, a single breather already has a larger
  $\lambda_2$ so that either one- or two-breather solutions dominate
  the biased measure depending on the value of $\alpha_1$ and
  $\alpha_2$. In practice, to maintain a small cloning rate and
  nevertheless be sure of seeing two breathers, we used $\alpha_1=0$
  and increased $\alpha_2$.

  The question as to how one transitions from typical profiles
  (homogeneous for $\alpha_i=0$) to atypical ones (with breathers for
  large enough $\alpha_i$) is an interesting one. In particular, it
  would be interesting to know whether there is a phase transition for
  critical values $\alpha_i^c$ and if so how the $\alpha_i^c$s scale with
  the system size. This is left for future work and we shall only here
  prove that the method can indeed single out atypical trajectories in
  spatially-extended systems.}

\begin{figure}
  \begin{tabular}{c}
    {\includegraphics{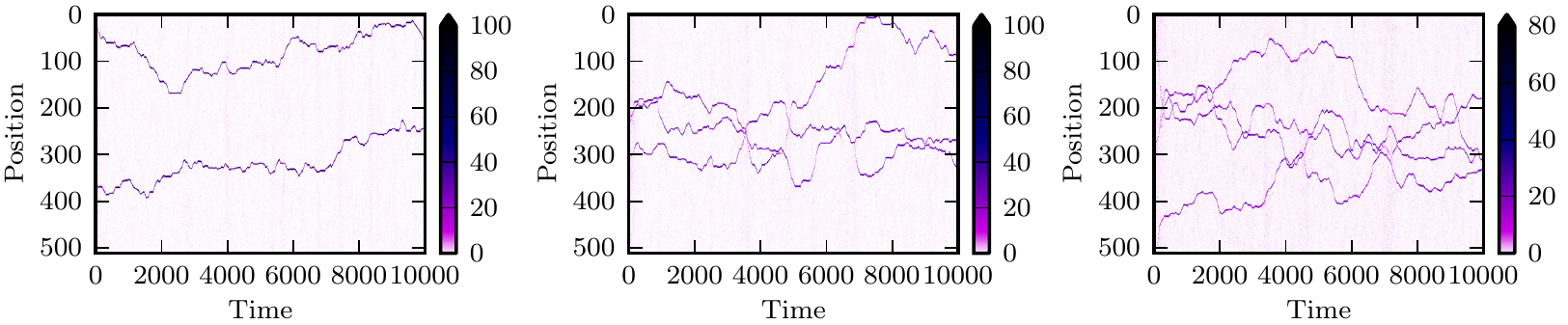}}
  \end{tabular}
  \caption{Multi-breather solutions in a $L=512$ system obtained using
    LWD with 200 clones and parameters $\dt=10^{-2}$ and
    $\eps=10^{-2}$. {\bf Left}: $\alpha_{i\neq 2}=0$, $\alpha_2=5L$
    yields two breathers. {\bf Center}: $\alpha_{i\neq 3}=0$,
    $\alpha_3=5L$ yields three breathers. {\bf Right}:  $\alpha_{i\neq 4}=0$,
    $\alpha_4=5L$ yields four breathers.}
  \label{FPU2b}
\end{figure}

Interestingly, it was reported in~\cite{Ruffo1} that the formation of
a single breather starting from a short wavelength perturbation of the
homogeneous state relied on the merging of breathers upon
collisions. The only way to observe a multi-breather solution thus
seemed to have a system large enough that two breathers would not have
the time to meet during their life time. If that was the case, the
solution presented in figure~\ref{FPU2b} would thus be artefact due to
noise and cloning. To check that this is not the case, we pursued the
simulations without noise and cloning, and found that  the
breathers indeed remain (meta)stable, even after encounters. This revealed a
great diversity in breather-breather interactions, as plotted in
figure~\ref{collision}. In addition to the coalescence reported
in~\cite{Ruffo1}, the breather-breather interaction can also be
repulsive, the breathers turning back before colliding, or they can
simply cross each other with little energy exchange. 

{To conclude, multi-breather solutions thus exist and were not detected
previously in the literature due to the difficulty of finding the
proper initial conditions. This is here unessential thanks to the
clone dynamics which automatically locate these highly aptypical
trajectories~\footnote{We thank the anonymous referee for suggesting
  this synthetic summary.}.}

\begin{figure}
  \includegraphics{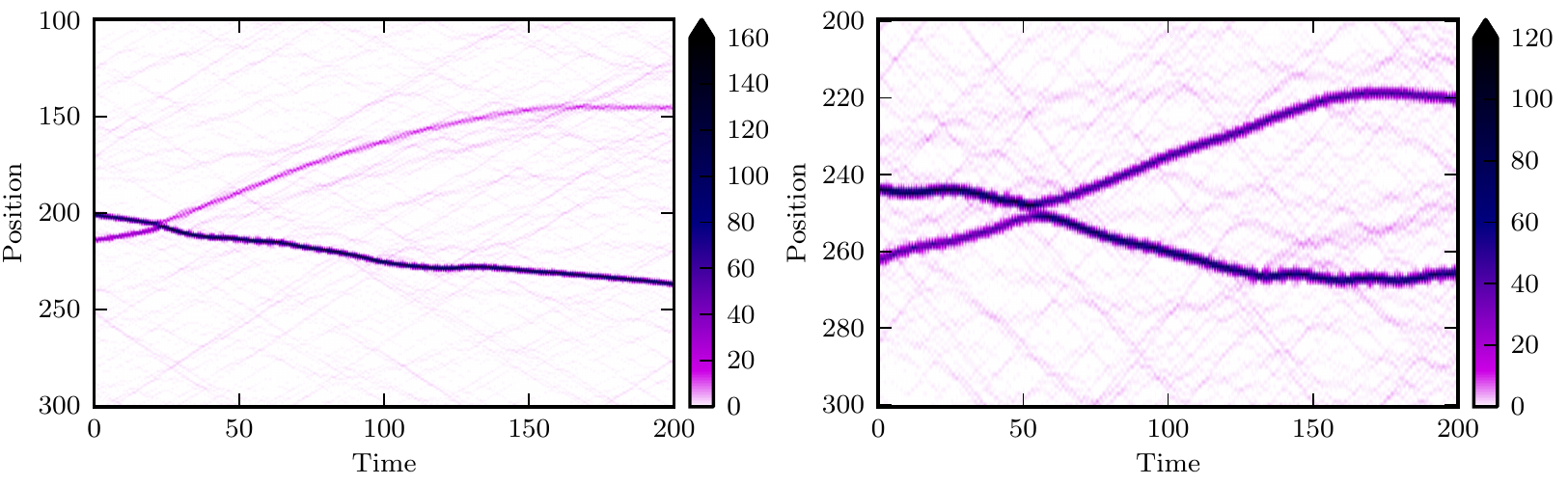}
  \caption{Close-up on two types of breather-breather interactions:
    crossing and repulsion, from left to right. We used the position
    and impulsion of a given clone as initial conditions for standard
    Hamiltonian dynamics.}
\label{collision}
\end{figure}

\if{As pointed out in REF, the Gram-Schmidt vectors are not eigenvectors of
$U^\dagger U$ and as such they contain a limited physical
information. It is nevertheless interesting to notice that in
situation with $n$ breathers, the $n$ first tangent vectors are
localized on the breathers but not in a one to one fashion. We
actually hit there a limitation of our approach: if all the breathers
are equivalent, then the spectrum becomes degenerate. The tangent
vectors can then ``swap'' rank and $v_3$ can become associated with the
second largest growing direction, so that a 2-breathers solution
becomes more interesting that a 3-breathers one, which is less probable.

detail techniques: bruits, cloning}\fi

\subsection{Large deviation functions: the example of coupled maps}
\label{sec:LDFCTM}
Beyond the detection of rare trajectories, one may be interested
in measuring their actual probability of occurrence. Many
systems exhibit intermittency, and show an apparent metastability in
space-time of trajectories of different chaoticity levels. The construction
of the cumulant generating function  gives access to a free-energy
like quantity that allows one to export the language of phase transitions
to such dynamical phase coexistence~\cite{Ruelle}. For instance, the
average Lyapunov exponent $\lambda_1(\alpha)$ in the clone population
plays the role of an order parameter since it is the first derivative
of the dynamical free energy $\mu(\alpha)$:
\begin{equation}
  \lambda_1(\alpha) = \frac{\langle \lambda_1 e^{\alpha \lambda_1 t}\rangle}{\langle e^{\alpha \lambda_1 t}\rangle}=\mu'(\alpha)
\end{equation}

As far as we know, the cumulant generating function of Lyapunov
exponents has only be computed for low dimensional models, such as
maps of the interval~\cite{bohr}, with the notable exception of the
Lorentz gas~\cite{LG1,LG2,LG3}. Recently, Vanneste extended the
algorithms introduced in~\cite{Tailleur2007} to compute the generating
function of the Lyapunov exponents of products of random
matrices~\cite{Vanneste}. He showed how the cloning method helps
reducing the variance of estimators based on finite number of clones
and computed the large deviation functions for products of small
random matrices (up to $8\times 8$). Here we show how the LWD can be
used to compute the cumulant generating function of a spatially
extended dynamical system: a one-dimensional lattice of coupled skewed
tent maps (CSTM)~\cite{Ginelli2007}.

\begin{figure}[h]
  \begin{center}
    \includegraphics{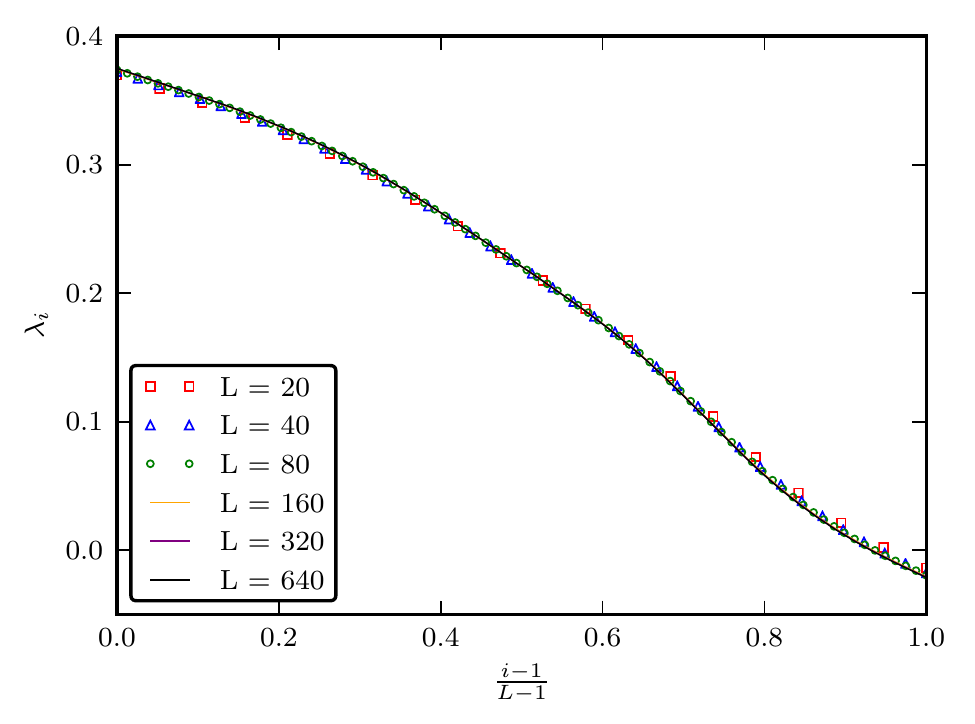}
  \end{center}
  \caption{Spectrum of Lyapunov exponents for
    $L=20,\,40,\,80,\,160,\,320$ and $640$. The exponents are averaged
    over a large number of runs, from $5\,000$ for $L=640$ to $500\,000$
    for $L=20$. For practical purpose, the spectrum has converged when
    $L\geq 40$.}
  \label{fig:spectrum}
\end{figure}

We consider $L$ real variables $x_i\in[0,1]$ whose discrete-time dynamics are given by
\begin{equation}\label{eqn:CTM}
  x_i(t+1) = f(x_i(t)) + D [f(x_{i+1}(t))+f(x_{i-1}(t))-2f(x_i(t))]\qquad 
  f(x)= \cases{b x \qquad \text{if } x\leq
    \frac{1}b\\ \frac{1-x}{1-\frac 1b} \quad  \text{otherwise}}
\end{equation}
When $D=0$, all the maps are uncoupled and fully chaotic. The
stationary measure is uniform and all Lyapunov exponents are equal
to~\cite{Ott}
\begin{equation}
  \lambda=\int_{0}^{1} \log|f'(x)| \d x=\frac 1 b \log b +
  \left(1-\frac 1 b\right) \log\frac{b}{b-1}
\end{equation}
When $D$ is increased, the maps become locally coupled and the
dynamics are made more regular by the diffusion terms, whence a
decrease of the Lyapunov exponents (we use $D=0.1$ and $b=4$ in the
following). This system is a good test for our method because it is
simpler to simulate than continuous time dynamics and its Lyapunov
spectrum converges quite rapidly when $L$ increases (see
figure~\ref{fig:spectrum}), compared to more complicated systems such
as the FPU chain of section~\ref{sec:FPU}. Furthermore, we may check
whether our algorithm allows us to go beyond the existing works on the
fluctuations of Lyapunov exponents in coupled-map lattices, which have
been limited to the study of the Gaussian regime~\cite{Kuptsov2011}.

To compute the Lyapunov exponents, we need to linearize the
dynamics~\eqref{eqn:CTM} to determine the time evolution of a tangent
vectors $\bol u$:
\begin{eqnarray}
  u_i(t+1) &=& (1-2 D) f'(x_i(t))u_i(t) + D [f'(x_{i+1})u_{i+1}(t)+f'(x_{i-1})u_{i-1}(t)]\\[2mm]
  f'(x)&=& \cases{b \qquad \text{if } x\leq
    \frac{1}b\\ \frac{b}{1-b} \quad  \text{otherwise}}
\end{eqnarray}
Again, to use the LWD we will need to make~\eqref{eqn:CTM}
stochastic. To do so, we use the shift
\begin{equation}
  \label{eqn:noise}
  \tilde x_i(t) = x_i(t) + \frac {\eps U}2 \times \text{min}(x_i(t),1-x_i(t))
\end{equation}
where $U$ is a random number sampled uniformly in $[-1,1]$, $\eps$ set
the scale of the noise and $\text{min}(x_i,1-x_i)$ makes sure $\tilde
x_i(t)$ remains in $[0,1]$ for $\eps<2$. We then replace $x_i(t)$ by
$\tilde x_i(t)$ in~\eqref{eqn:CTM}. {The
  ``noise''~\eqref{eqn:noise} is clearly not the Gaussian white noise
  discussed in section~\ref{sec:method} for Hamiltonian dynamics. One
  thus has to be careful and check that the observed fluctuations of
  Lyapunov exponents do not depend on the details and intensity of the
  noise (see below).}

We first use brute-force simulations for $L=40$ to estimate the
Gaussian scaling of the fluctuations of $\lambda_1$ and check the
effect of the noise~\eqref{eqn:noise} on the fluctuations of the
coupled maps. After an initial run of 500 time-steps, the Lyapunov
exponents are computed over the next {$t=10^4$} iterations of the
maps. For $N=10^6$ simulations, figure~\ref{fig:bruteforce} shows
that the pdf is well approximated by a Gaussian $P_G(\lambda_1,t)$:
\begin{equation}
 {P_{\rm sim.}(\lambda_1,t)\simeq P_G(\lambda_1,t)\equiv \sqrt{\frac t {2\pi\sigma^2}} e^{-\frac t 2 \frac{(\lambda_1-\langle
      \lambda_1\rangle)^2}{\sigma^2}}}\quad\text{with}\quad \langle
  \lambda_1\rangle \simeq 0.372 \quad  \sigma^2 \simeq
  2.03\times 10^{-2}
\end{equation}

Note that the standard deviation of $\lambda_1$ is
$\tilde\sigma=\sigma/\sqrt{t}\sim 4.\, 10^{-3} \langle
\lambda_1\rangle $, showing that these simulations do not allow to
sample very far from the average value. Nevertheless,
figure~\ref{fig:bruteforce} shows the beginning of the large
deviations regime since there seems to be a small but systematic
asymmetric deviations from the Gaussian approximation when
$P(\lambda_1,t)\sim 10^{-4}-10^{-6}$. Importantly, simulations with
levels of noise $\eps=10^{-2}$ and $\eps=10^{-4}$ show no difference
in $P(\lambda_1,t)$, so that $\eps=10^{-2}$ will thus be used in the
following.
\begin{figure}[h]
  \begin{center}
\includegraphics{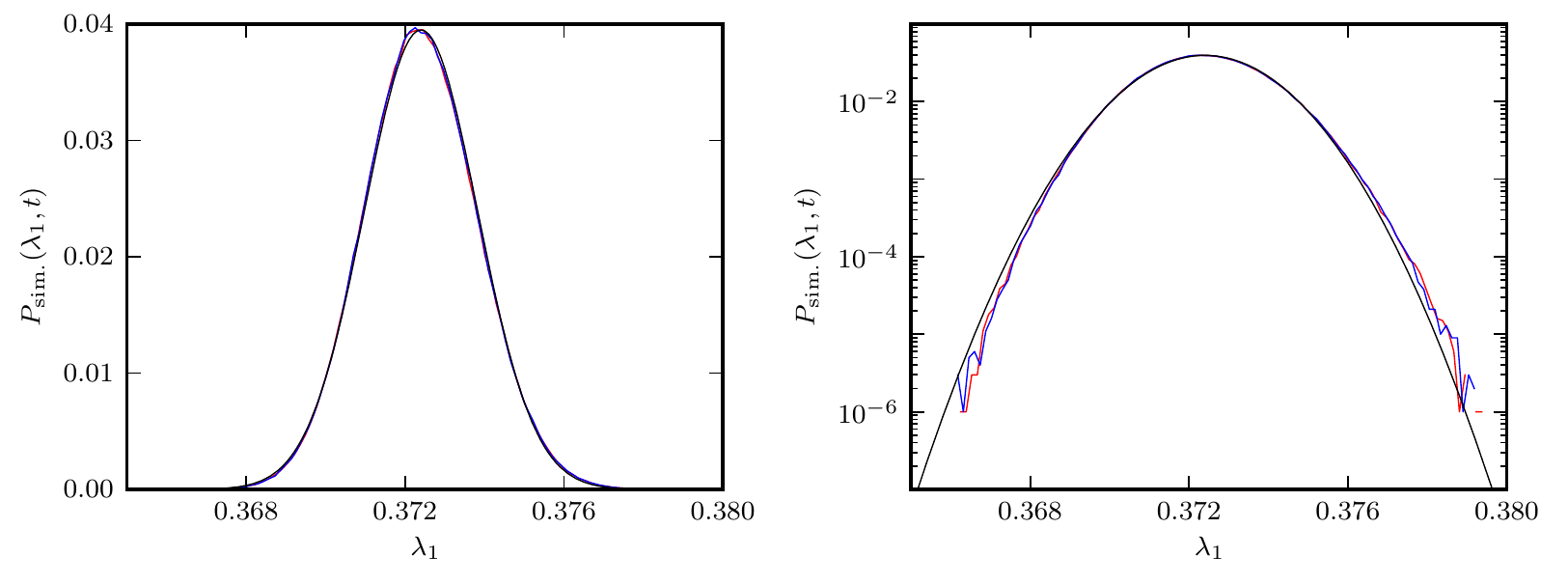}
  \end{center}
  \caption{{\bf Left}: $P_{\rm sim.}(\lambda_1,t)$ constructed from direct sampling
    of $10^6$ simulations for $L=40$, $D=0.1$, $b=4$, $\eps=10^{-2}$
    (red) or $\eps=10^{-4}$ (blue). The black line is the Gaussian
    approximation. {\bf Right}: same plot in logscale. One sees a
    small asymmetric deviation from the Gaussian when
    $P(\lambda_1,t)\sim 10^{-4}-10^{-6}$.}
  \label{fig:bruteforce}
\end{figure}

From the Gaussian approximation, one can construct a quadratic
approximation to $\mu(\alpha)$
\begin{equation}
  \mu^G(\alpha)=\frac 1 t \log \Big[\int_{-\infty}^{\infty} {\rm e}^{\alpha \lambda_1 t}
  P_G(\lambda_1,t) \d \lambda_1\Big] = \alpha \langle \lambda_1 \rangle +\frac 1 2 \alpha^2
  \sigma^2
\end{equation}
Similarly, the value of the Lyapunov exponent $\lambda_1(\alpha)$ that
dominates the biased measure for a fixed value of $\alpha$ is
approximated by
\begin{equation}
  \lambda_1^G(\alpha)=\frac{\langle \lambda_1 {\rm e}^{\alpha
      \lambda_1 t}\rangle}{\langle{\rm e}^{\alpha \lambda_1
      t}\rangle}=\partial_\alpha \mu^G(\alpha,t)=\langle \lambda_1 \rangle + \alpha \sigma^2
\end{equation}

One can then compare the Gaussian approximations with $\mu(\alpha)$
and $\lambda_1(\alpha)$ obtained using the LWD of
section~\ref{sec:LWD_Algo}.  {The simulations in
  figure~\ref{fig:fullldf} were obtained using varying numbers of
  clones, from $N_c=100$ to $N_c=160\,000$. For each value of $\alpha$
  and $N_c$, $\lambda_1(\alpha)$ and $\mu(\alpha)$ were obtained by
  averaging over 10 runs of $N_c$ clones. For $\alpha<0$ the
  convergence with the number of clones is very fast, the difference
  in $\lambda_1(\alpha)$ between $N_c=100$ and $N_c=160\,000$ for
  $\alpha=-3$ is less than $0.5\%$. For $\alpha>0$, the convergence
  with $N_c$ is slower when $\alpha$ gets larger. To estimate the
  speed of convergence, one can check for which value of $\alpha$ the
  difference between the estimators of $\lambda_1(\alpha)$ for $N_c$
  and $160\,000$ clones goes above $5\%$. One gets
  $\alpha=1.4,\,2.0,\,2.2,\,2.4$ for $N_c=100,\,500,\,1\,000$ and
  $N_c=2\,000$. The estimators for $N_c=10\,000$, $20\,000$, $40\,000$
  and $80\,000$ always remain within $3.5\%$, $2\%$, $1\%$ and $0.5\%$
  of the values obtained for $N_c=160\,000$. The same simulations done
  with different noise intensities ($\eps=10^{-3}$ or $\eps=5.\,
  10^{-2}$) lead to the same asymptotic values for $\lambda_1(\alpha)$
  and $\mu(\alpha)$ when $N_c\to\infty$, though the convergence rate
  is slower at smaller noise intensities (see
  figure~\ref{fig:noiseconv}). For $N_c=160\,000$, the difference
  between $\lambda_1(\alpha)$ for the three noise intensities is
  smaller than $0.5\%$ for $\alpha\leq 2.2 $ whereas it goes up to
  $2\%$ at $\alpha=3.0$.}

{Let us highlight here that the number of clones used to compute
  $\lambda_1(\alpha)$ is much larger than what was used in
  section~\ref{sec:FPU} to detect the breathers. This is mostly due to
  the fact that in section~\ref{sec:FPU}, we simply wanted to detect
  an atypical trajectory, without quantifying its rarity, whereas in
  this section we aim at computing averages that require sampling over
  many different trajectories. There is no obvious rule as to when a
  large number of clones will be needed to compute $\lambda_1(\alpha)$
  (for $\alpha<0$ the convergence is very fast, for instance). Also,
  many tricks exist in the literature for Sequential Monte Carlo
  simulations (see~\cite{DelMoral2012} and reference therein) that
  could/should probably be extended to the LWD to speed up the
  convergence of the algorithm.}

\begin{figure}
\includegraphics{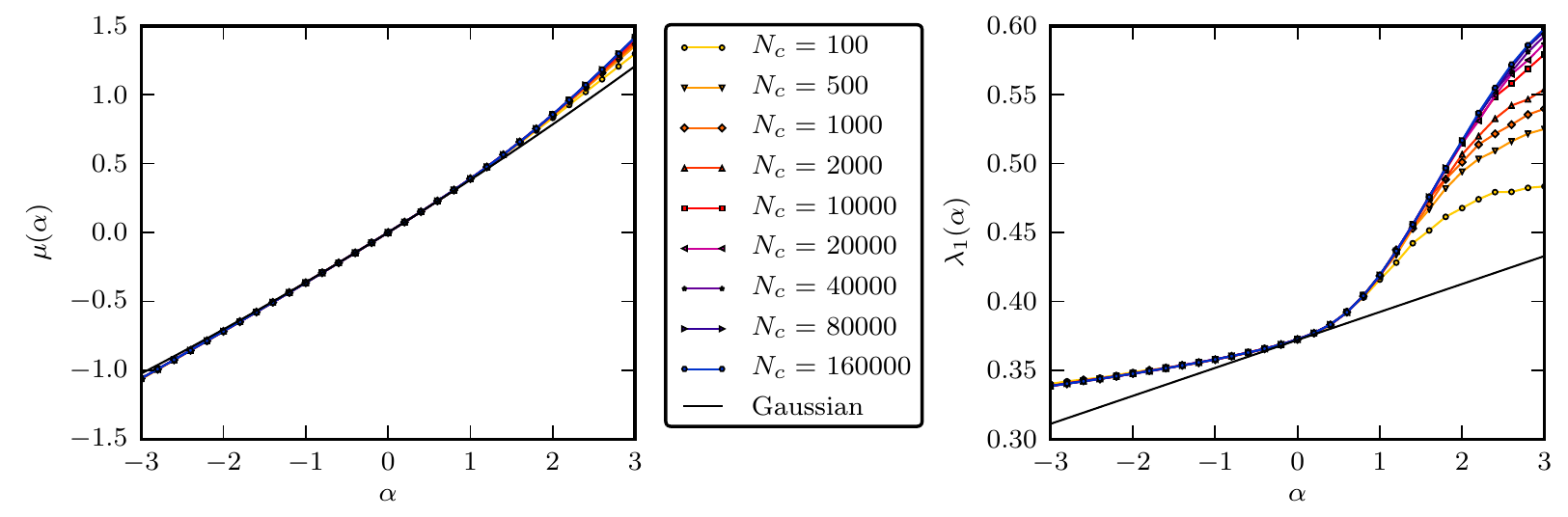}
  \caption{$\mu(\alpha)$ (left) and $\lambda_1(\alpha)$ (right) for
    $L=40$ using $N_c$ clones where $N_c$ goes from $100$ to $160\,000$. The
    solid black lines correspond to the Gaussian approximations, valid
    close to $\alpha\simeq 0$, whereas the symbols come from the
    cloning simulations.} 
  \label{fig:fullldf}
\end{figure}

\begin{figure}
  \begin{center}
    \includegraphics{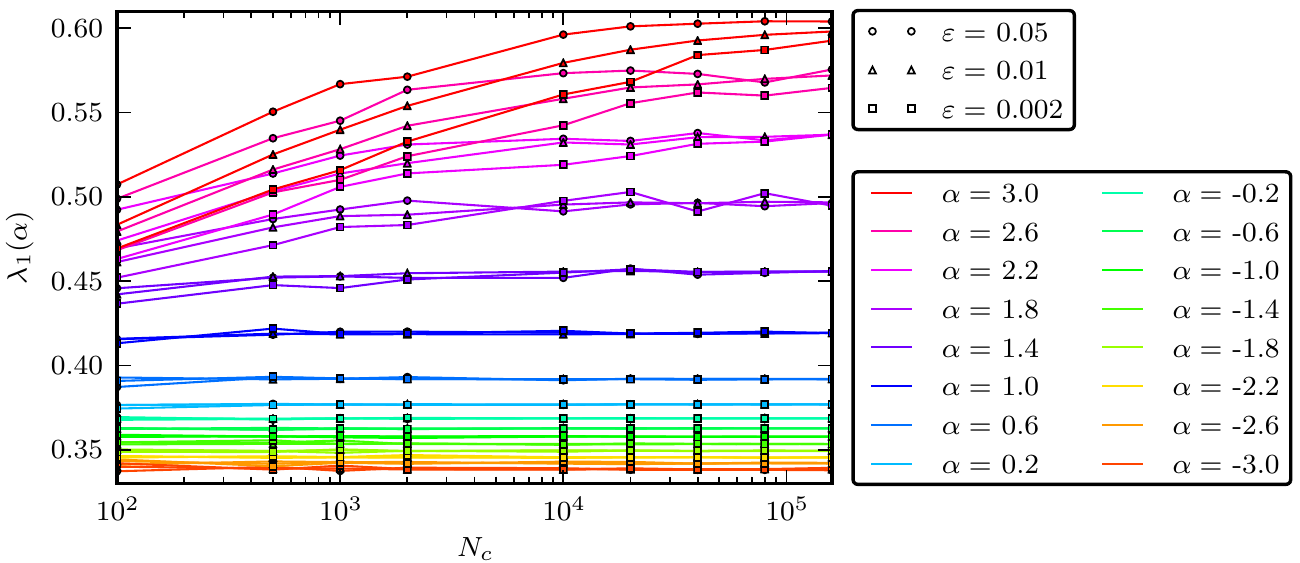}
  \end{center}
  \caption{Convergence of $ \lambda_1(\alpha)$ as a
    function of the clone number $N_c$ for three levels of noise. A
    given color corresponds to a given value of $\alpha$ while
    different symbols correspond to different levels of noise.}
  \label{fig:noiseconv}
\end{figure}

In figure~\ref{fig:fullldf} we see that the LWD indeed samples far
beyond the Gaussian regime: the values of $\lambda_1(\alpha)$ vary
over an interval of order $\langle \lambda_1 \rangle$ whereas $\tilde
\sigma$ was less than $1\%$ of $\langle \lambda_1\rangle $. Let us
also note that the deviations from the Gaussian scaling are more clearly
seen in $\lambda_1(\alpha)$ than in $\mu(\alpha)$. This can be
understood because the latter is nothing but the integral of the
former. Since $\lambda(\alpha)$ is quite small ($\sim 0.5$ here), we
need a large interval of $\alpha$ for the error on $\lambda_1(\alpha)$
to have an impact on $\mu(\alpha)$. Nevertheless, it is quite
reassuring that our simulations agree with the Gaussian scaling when
$\alpha\simeq 0$.

Since there are neither analytical nor numerical benchmarks for our
computations of $\mu(\alpha)$ and $\lambda_1(\alpha)$, we compared the
results obtained using two different types of cloning
algorithms. {We used both the methods described in the core and at
  the end of section~\ref{sec:LWD_Algo}, that use resampling of the
  clone population respectively clone-by-clone or globally}. Both
methods gave similar results, the clone-by-clone method (which we
generically refer to when we speak about LWD) being substantially
faster that the global resampling scheme. We have thus shown that LWD
can be used in spatially extended systems to compute the cumulant
generating functions beyond the Gaussian regime. This now opens the
way to systematic studies of the dependence of $\mu(\alpha)$ and
$\lambda_1(\alpha)$ with the system size, which should allow us
to discuss the possibility of dynamical phase transitions separating
different scalings with the system size of the exponents fluctuations.

\section{Method}
\label{sec:method}
\subsection{Definition of finite-time Lyapunov exponents}
Let us consider a system evolving under Hamiltonian dynamics, which can
be written in a concise form
\begin{equation}
  \label{eqn:hameq}
  \dot x_i = f_i[{\bf x}(t)]\qquad\mbox{with}\qquad
  \cases{{\bf x}=(q_1,\dots,q_N,p_1,\dots,p_N)\\
    {\bf f}=(\frac{\partial \H}{\partial p_1},\dots,\frac{\partial \H}{\partial p_N},-\frac{\partial \H}{\partial {q_1}},\dots,-\frac{\partial \H}{\partial {q_N}})}
\end{equation}
The Lyapunov exponents characterize the exponential convergence or
divergence between two nearby trajectories. They are usually defined
by considering the linearized dynamics of a tangent vectors ${\bol
  \omega}$:
\begin{equation}
  \label{tangdyn}
   \dot {\bol \omega} = -\mathbb A  {\bol \omega}\qquad\text{with}\qquad
  \mathbb A_{ij}=-\frac{\partial f_i[\bol x(t)]}{\partial x_j}
\end{equation}

There are two ways of defining finite-time Lyapunov exponents. For a
given trajectory, let us call $U(t)$ the matrix solution of the linear
equation $\dot U(t)=-\mathbb A[{\bf x(t)}] U(t)$ and ${\cal A}_1(t)\geq{\cal
  A}_2(t)\dots\geq {\cal A}_N(t)$ the $N$ first eigenvalues of $U^\dagger U$. We
then obtain a first definition of the finite-time Lyapunov exponents as
\begin{equation}
  \label{expm}
  \lambda_i(t)= \frac 1 {2t} \log{\cal A}_i(t)
\end{equation}
Alternatively, one can look at a $k$-volume $V_k(t)={\bol \omega}_1(t)\wedge {\bol
  \omega}_2(t) \dots\wedge {\bol\omega}_k(t)$ and define
\begin{equation}
  \label{expv}
  \lambda_k(t)=\lim_{t\to\infty}\frac 1 t \log|{V}_k(t)|-\lim_{t\to\infty} \frac 1 t\log|{V}_{k-1}(t)|
\end{equation}
which thus corresponds to the exponential growth in the $k$th
dimension of the $k$-volume $V_k$. In the limit $t\to\infty$, for
generic choices of the vectors $\bol \omega_i$, both definitions
coincide, although they generically differ at finite time. Expression
\eqref{expm} is more satisfactory because the exponents do not depend
on the choice of the ${\bol \omega}_i$s, but \eqref{expv} is much
simpler to implement numerically. The method developed in this paper
uses a Gram-Schmidt orthogonalization procedure that relies on
definition~\eqref{expv}, but it would be interesting to see how the
\textit{fluctuations} of finite-time Lyapunov exponents differ using
definition~\eqref{expm}. A natural way to do so may be to use the
recent algorithms introduced to compute covariant Lyapunov
vectors~\cite{Ginelli2007,Kaz2011}.

{Let us note furthermore that both \eqref{expm} and \eqref{expv}
  explicitly depend on the choice of trajectory $x(t)$ around which
  the linearized dynamics~\eqref{tangdyn} is studied, and hence we
  could/should write $\lambda_i[t,x(0)]$. For non-ergodic systems,
  even the long time limit $\lambda_i(t\to\infty)$ still depends on
  $x(0)$ whereas for ergodic systems Osedelets ergodic multiplicative
  theorem ensures that $\lambda_i(t\to\infty)$ adopts the same value
  for (almost all) $x_0$~\cite{beck}. In the following we drop the
  dependence on $x_0$ but one should remember that it is an important
  source of fluctuations of $\lambda_i(t)$.}

Let us show how to relate equation~\eqref{expv} to the tangent
dynamics~\eqref{tangdyn}.  To do so, starting from the vector ${\bol
  \omega}_i$, we obtain a set of orthogonal vectors ${\bol u}_i$
which generate the same oriented volume $V_k$ through
\begin{equation}
  \bol u_i= \bol \omega_i - \sum_{j=1}^{i-1} \bol u_j \frac{\bol u_j
    \cdot \bol \omega_i}{\bol u_j \cdot \bol u_j} 
\end{equation}
whose dynamics can be inferred from~\eqref{tangdyn}
\begin{equation}
  \dot \bol u_i = -\mathbb A \bol u_i + \sum_{j=1}^{i-1} \bol u_j
  \frac{\bol u_i \cdot \mathbb A \bol u_j + \bol u_j \cdot \mathbb A \bol u_i}{|\bol u_j|^2}
\end{equation}
One directly checks that $\frac{\d}{\d t} (\bol u_i \cdot \bol
u_j)=0$ so that the $\bol u_i$s remain orthogonal if they were chosen so at
$t=0$. Furthermore, $\bol \omega_1\wedge\bol \omega_2 \dots\wedge
{\bol\omega}_k =\bol u_1 \wedge \bol u_2 \dots\wedge \bol u_k$, since
an exterior product involving twice the same vector vanishes.

Let us then introduce the inner product
\begin{equation}
  \langle  \bol u_1
\wedge \bol u_2 \dots\wedge \bol u_k | \bol v_1
\wedge \bol v_2 \dots\wedge \bol v_k \rangle = \det M
\qquad\text{with}\qquad M_{ij} = \bol u_i \cdot \bol v_j
\end{equation}
to express the norm of $V_k$ through $|V_k|^2= \langle \bol u_1(t)
\wedge \bol u_2(t) \dots\wedge \bol u_k(t) | \bol u_1(t) \wedge \bol
u_2(t) \dots\wedge \bol u_k(t) \rangle$ so that the time evolution of
$V_k$ is given by
\begin{equation}
\frac{\d}{\dt} |V_k|^2 =  2 \sum_j \langle \bol u_1 \wedge \dots \wedge \dot {\bol u}_j \wedge \dots \wedge \bol u_k | \bol u_1 \wedge \dots \bol u_k \rangle
\end{equation}
All terms in $ \dot{\bf u}_j$ parallel to $\bol u_{i\neq j}$ vanish, yielding
\begin{equation}
\frac{\d}{\dt} |V_k|^2 = - 2 \sum_j \langle \bol u_1 \wedge \dots \wedge \mathbb A  {\bol u}_j \wedge \dots \wedge \bol u_k | \bol u_1 \wedge \dots \bol u_k \rangle
\end{equation}
We thus have to compute a sum of determinants of the type
\begin{equation*}
  \left|\matrix{
    |\bol u_1|^2 &0&\dots&&\dots&0\cr
    0&|\bol u_2|^2&0&&\dots&0\cr
    \vdots&\vdots&\ddots&&&\vdots\cr
    \bol u_1 \cdot \mathbb A \bol  u_j&\bol u_2\cdot \mathbb A\bol u_j&\dots&\bol u_j\cdot \mathbb A\bol u_j&\dots&\bol u_k \cdot \mathbb A \bol u_k\cr
    \vdots&\vdots&\dots&&&0\cr
    0&\dots&&&\dots&|\bol  u_k|^2
  }\right|=|\bol  u_1|^2\dots |\bol u_{j-1}|^2 \bol u_j\cdot  \mathbb A \bol u_j |\bol u_{j+1}|^2 \dots |\bol u_k|^2
\end{equation*}
and thus
\begin{equation}
  \label{eqn:dynVk}
  \ddt{} |V_k|^2 = -2 \prod_{i=1}^k |\bol  u_i|^2 \sum_{i=1}^k \bol
  v_i\cdot \mathbb A \bol v_i = -2 |V_k|^2 \sum_{i=1}^k \bol v_i \cdot \mathbb A \bol v_i
\end{equation}
where we have introduced the unitary vectors $\bol  v_i=\frac{\bol u_i}{|\bol
  u_i|}$ whose evolution read
\begin{equation}
\label{eqn:evolvi}
  \dot {\bol v}_i = -\mathbb A \bol v_i + \bol v_i (\bol v_i \cdot \mathbb A \bol v_i) + \sum_{j=1}^{i-1} \bol v_j (\bol v_i \cdot \mathbb A \bol v_j + \bol v_j\cdot \mathbb A \bol v_i)
\end{equation}
Finally \eqref{eqn:dynVk} can be solved to yield
\begin{equation}
  |V_k(t)| =\ee^{- \int_0^t \dt \left\{\sum_{i=1}^k \bol v_i \cdot \mathbb A \bol v_i\right\}} |V_k(0)|
\end{equation}
One can thus rewrite~\eqref{expv} as
\begin{equation}
  \label{eqn:deflambdakt}
  \lambda_k(t)=-\frac 1 t \int_0^t \dt \left\{\bol  v_k\cdot  \mathbb A \bol  v_k\right\}
\end{equation}
This formalism may seem complicated but~\eqref{eqn:evolvi} simply
yields the continuous time evolution of the Gram-Schmidt
vectors~\cite{Wiesel93}, while the $\lambda_k(t)$ defined
in~\eqref{eqn:deflambdakt} can be computed using the rescaling factors
that appear in the renormalization steps of the Gram-Schmidt
procedure. This procedure indeed corresponds to evolving $k$ unitary
vectors $\bol v_i$ initially chosen orthogonal with the tangent
dynamics
\begin{equation}
  \label{eqn:nosotan}
\dot {\bol v}_i = -\mathbb A \bol v_i
\end{equation}
At every time step $t=r\dt$ we re-orthogonalize them and normalize them,
calling {$s_j(r\dt)$} the rescaling factor of $\bol v_j$. Then, when $\dt\to 0$, the
dynamics of $\bol v_i$ tends toward
(\ref{eqn:evolvi}) and
\begin{equation}
  {\prod_{r=1}^{t/\dt}\;  \prod_{j=1}^k \;} {s_j(r\dt)}\simeq\exp \Big(-\int_0^t\dt \sum_{j=1}^k \bol v_j \cdot \mathbb A \bol v_j\Big)= \exp \Big( t \sum_{j=1}^k \lambda_j(t)\Big)
\end{equation}
Let us now show how this formalism allows us to sample the large
deviations of Lyapunov exponents.

\subsection{Large deviation formalism}

In general, one needs different levels of
control of the details of the algorithm depending on whether one wants to compute a dynamical free energy (or
topological pressure) or simply look for atypical trajectories.
 One important
point is that for the simulations, it is always necessary  to add a small amount  of
noise to the system. We shall assume that this is the case below, 
but  one should remember that in some cases
the noise is simply an artifact and has to be chosen as small as
possible. We thus consider a slightly different version of the
dynamics~\eqref{eqn:hameq}:
\begin{equation}
\label{eqn:hamequabruit}
\dot q_i = \ddp{\H}{p_i}\qquad \dot p_i = -\ddp{\H}{q_i} + \sqrt{2 \eps} \eta_i
\end{equation}
where $\eta_i$ are unitary Gaussian white noises. To compute the
dynamical partition function
\begin{equation}
  \label{eqn:Z}
  Z(\alpha,t)=\langle e^{\alpha \lambda_1  t}\rangle
\end{equation}
we first need to define the average $\langle \,\dots\,\rangle$. In
deterministic dynamical systems, finding a putative stationary
measure is difficult and the addition of a Gaussian white noise
simplifies drastically the situation. For
instance,~\eqref{eqn:hamequabruit} would lead to a flat measure over
the whole phase space,  while considering a stochastic force acting tangentially to the
energy surface would make the microcanonical measure be the correct
steady-state (see~\ref{sec:bruitcons}).  In the latter case, the
average~\eqref{eqn:Z} then amounts to choosing the initial conditions
uniformly over the energy surface and averaging over the noise
realizations. 

From the mathematical point of view, the addition of a
small noise changes the nature of the system. The question as to
whether one recovers, in the small noise limit, the steady-state measure
of the underlying deterministic system dates back to Kolmogorov
(see~\cite{Sinai89}). This question of ``stochastic
stability''~\cite{Young05} is actually a natural way for a physicist
to define the steady-state measure of a dynamical system, and it is
thus fortunate that when SRB measures exist, they can indeed be recovered
as small noise limits of stochastic dynamical systems
(see~\cite{Young05} for a proper mathematical presentation of this
procedure).

\subsubsection{Largest Lyapunov exponent}
For simplicity, we first consider the large deviations of the largest
Lyapunov exponent $\lambda_1$ and later generalize the formalism to
fluctuations of several exponents. Furthermore, we will use the
Gaussian noises introduced in equation~\eqref{eqn:hamequabruit} and
discuss in~\ref{sec:bruitcons} how to proceed with conserved
noises. The probability density obeys a modified Liouville equation~\footnote{Derivatives apply to everything on their
  right when they appear as operator, as in  $\ddp{}{x}$, and not when explicitly applied to a function, as in $\ddp{f}{x}$.}:
\begin{equation}
  \label{eqn:hameps}
  \ddp{P(\bol q,\bol p,t)}{t}= -H_\eps P(\bol q,\bol p,t)\qquad H_\eps=\sum_{i=1}^N -\eps \ddp{^2}{p_i^2} +\ddp{\H}{p_i} \ddp{}{q_i} - \ddp{\H}{q_i}\ddp{}{p_i}
\end{equation}
Including the dynamics of the tangent vector $\bol v$, {and using the notation $\bol x=(\bol q,\bol p)$}, this becomes
\begin{equation}
  \label{eqn:hamepsprime}
  \ddp{P(\bol x,\bol v,t)}{t}= -H P(\bol x,\bol v,t)=-\Big(H_\eps - \sum_{i=1}^{2N} \ddp{}{v_i}\big[\sum_j \mathbb A_{ij} v_j +N(\bol v)v_i\big]\Big)P(\bol x,\bol v,t)
\end{equation}
where we have introduced $N(\bol v)=-\sum_{ij=1}^{2N} v_i \mathbb A_{ij} v_j$ for
clarity. We can then define a distribution
of $\lambda_1(t)$ through
\begin{equation}
  P(\lambda_1,t) = \int {\cal D}[\bol x,\bol v] \d\bol x_0\d\bol v_{0}  \delta[\lambda_1-\lambda_1(\bol x,\bol v,t)]P(\bol x_0,\bol v_{_0})
\end{equation}
where $\lambda_1(\bol x,\bol v,t)$ is defined in
(\ref{eqn:deflambdakt}). The evolution of the unitary tangent vectors
$\bol v$ is given in \eqref{eqn:evolvi}.

We will now rewrite (\ref{eqn:Z}) as the path integral of a
generalized evolution operator in which we will then read the
algorithm  used to bias the trajectories with a weight
$\exp(\alpha \lambda_1 t)$. We wish to compute
\begin{equation}
  \langle e^{\alpha \lambda_1 t} \rangle = 
  \langle e^{-\alpha\sum_{ij}\int_0^t v_i \mathbb A_{ij} v_j \dt } \rangle
\end{equation}
which can be rewritten as
\begin{eqnarray}
    \langle e^{\alpha \lambda_1 t} \rangle &=
    \!\!\int\!\! {\cal D}[\bol x,\bol v ,\bol \eta]\,\d\bol x_0 \d\bol v_0 P(\bol x_0,\bol v_0) \prod_{i=1}^N
    \delta\left(\!  \dot q_i - \ddp{\H}{p_i}\right)\delta\left(\! \dot
    p_i +\ddp{\H}{q_i} - \sqrt{2 \eps} \eta_i\right)\nonumber\\
    &\prod_{i=1}^{2N}\delta \Big(\dot v_i+\sum_j \mathbb A_{ij}v_j -v_i \sum_{kl} v_k \mathbb A_{kl}v_l
    \Big)\ee^{- \sum_{i=1}^N \int_0^t\dt \frac{\eta_i^2}{2}}\,\ee^{-\sum_{ij=1}^{2N} \int_0^t \dt
      \alpha v_i \mathbb A_{ij} v_j }
\end{eqnarray}
We thus integrate over all possible trajectories starting from all
possible initial conditions. The $\delta$ functions constrain these
trajectories to be solutions of the equations of motion and the
Gaussian weight is the probability of a given noise realization. Since
we will only use this path integral to recognize an evolution
operator, we do not need to worry too much about operator ordering and
time-discretization of multiplicative noises. Using the complex
representation of $\delta$ functions and doing explicitly the Gaussian
integral over the noise, we get:
\begin{eqnarray}
  \label{eqn:pathintegra1}
  \langle e^{\alpha \lambda_1 t} \rangle &=
  \!\!\int\!\! {\cal D}[\bol x,\bol v,\hat{\bol x},\hat{\bol v}]\,\d\bol x_0\d\bol v_0
  \exp\Big[\int \dt \big\{\sum_{i=1}^N \big[\hat q_i \dot q_i + \hat p_i
        \dot p_i + \eps \hat p_i^2 - \ddp{\H}{p_i} \hat q_i +
        \ddp{\H}{q_i} \hat p_i\big]\\
    &+ \sum_{i=1}^{2N} \big[\hat
        v_i \dot v_i +\hat v_i \big(\sum_{j=1}^{2N}\mathbb A_{ij} v_j + v_i N(\bol v)
        \big)\big]+\alpha N(\bol v) \big\}\Big] P(\bol x_0,\bol v_0)
\end{eqnarray}
where $\hat x,\hat v$ are imaginary fields. This path integral corresponds to the
matrix element
\begin{equation}
  \langle e^{\alpha \lambda t} \rangle = \langle - | e^{- t \big(H
    -\alpha N(\bol v)\big) } |P_0(\bol x_0,\bol v_0) \rangle
\end{equation}
where $H$ is given by
\begin{equation}
  H= H_\eps - \sum_{i=1}^{2N} \ddp{}{v_i}\Big[\sum_{j=1}^{2N} \mathbb A_{ij} v_j +N(\bol v)v_i\Big]
\end{equation}

The cumulant generating function $\mu(\alpha)=\frac 1 t \log
Z(\alpha,t)$ is then given by minus the smallest eigenvalue of $H-\alpha N(\bol v)$. The
evolution operator $H-\alpha N(\bol v)$ does not correspond to a standard Langevin
equation since it does not conserve the total probability. The
evolution equation
\begin{equation}
  \label{eqn:evoldensmarch}
  \dot{P}=-(H-\alpha N(\bol v)) P
\end{equation}
indeed implies that 
\begin{equation}
  \int \text{d} {\bf x} \text{d} 
  {\bf v} \dot P (\bol x,\bol v;t)=-\alpha \int
  \text{d}{\bf x} \text{d} {\bf v} \;\;  \sum_{i,j}v_i \mathbb A_{ij} v_j P(\bol x,\bol v;t)
\end{equation}
which does not generically vanish. This means that if one wishes to
represent $P(\bol x,\bol v,t)$ by a population of points in the space
$(\bol x,\bol v)$, the number of points is not conserved. Each copy
the system is thus replicated (or pruned) at a rate $\alpha N(\bol
v)$, i.e. $\alpha$ times the stretching rate of the tangent vector
${\bol u(t)}$. Before we turn to the numerical implementation of this
dynamics, let us show how the discussion above extends to the
fluctuations of several Lyapunov exponents.

\subsubsection{Large deviations of several Lyapunov exponents}
\label{sec:ldfmany}
Let us consider the joint fluctuations of the $L$ first Lyapunov
exponents. We thus wish to compute the generating function
\begin{equation}
Z(\bol \alpha,t)=  \left\langle \ee^{-\sum_{k=1}^L \alpha_k \int_0^t \bol  v_k \cdot \mathbb A \bol v_k} \right\rangle
\end{equation}
where $\bol\alpha=(\alpha_1,\dots,\alpha_L)$ is a vector of ``biases''
corresponding to each Lyapunov exponents.  Following a similar path as
in the previous subsection, we can write $Z$ as
\begin{equation}
Z(\bol \alpha,t)
  = \langle - | e^{- t \big(H
    -\sum_{k=1}^L \alpha_k N(\bol v_k)\big) } |P_0(\bol  q_0, \bol p_0,
\bol v_{1,0},\dots,\bol v_{L,0}) \rangle\sim \ee^{t \mu(\bol \alpha)}
\end{equation}
where $\mu(\bol \alpha)$ is now minus the smallest eigenvalue of the evolution
operator $H-\sum_{k=1}^L\alpha_k N(\bol v_k)$, where
\begin{eqnarray}
    H&= \sum_{i=1}^{N} \left\{-\eps \frac{\partial^2}{\partial p_i^2
      } +\ddp{\H}{p_i} \ddp{}{q_i}-\ddp{\H}{q_i}\ddp{}{p_i}\right\}\\
    &\quad +\sum_{i=1}^{L} \left\{-\ddp{}{\bol v_i} \cdot \left(\mathbb A \bol v_i - (\bol v_i\cdot \mathbb A \bol v_i) \bol v_i- \sum_{j=1}^{i-1} \bol v_j (\bol v_i\cdot \mathbb A \bol v_j + \bol v_j \cdot \mathbb A \bol v_i) \right)\right\}
\end{eqnarray}
\begin{equation}
{  N(\bol v_k)=-\bol v_k \cdot \mathbb A \bol v_k}
\end{equation}
We thus have to represent $P(\bol x,\bol v_1,\dots,\bol v_L;t)$ by a
population of walkers in the space $(\bol x,\bol v_1,\dots,\bol
v_L)$. The pruning or cloning rate is now given by the product of each
of the factors $\alpha_k N(\bol v_k)$, which are nothing but
$\alpha_k$ times the stretching rate of the vector $\bol v_k$.

\subsection{Algorithm: the Lyapunov Weighted Dynamics (LWD)}
\label{sec:LWD_Algo}
Let us now turn to the numerical implementation of the dynamics
generated by $H-\sum_k\alpha_k N(\bol v_k)$. To do so, we will use a
population dynamics that allows one to simulate dynamics that do not
conserve probability. This type of algorithm is closely related to
Sequential Monte Carlo dynamics or Diffusion Monte
Carlo~\cite{Anderson75,Kalos69,Ceperley79,Nightingale86,Nightingale88,Gelbard90,Martin95}. An
alternative to this type of approach is to carry Metropolis like
Monte Carlo directly in the trajectory space, as does
\textit{Transition path sampling}~\cite{Dellago98} which has been used
to compute large deviation functions of dynamical observables in
models of glass forming liquids~\cite{Hedges2009} (see~\cite{Geiger}
for a version of transition path sampling that uses Lyapunov exponents
to weigh the trajectories).

We consider a population of $N_c$ copies of the system with
positions and momenta $\bol q$ and $\bol p$. To each copy $j$, we
attach $L$ tangent vectors $\bol v_i$. We then choose a time-step
$\dt$ and evolve the system {over $T$ time steps so that the total
  time is $t=T\dt$}. For $t=0$, the ${N_c}$ copies start from
arbitrary initial conditions. At each time step $t'=n \dt$:
\begin{itemize}
\item[\fbox{$1$}] For each copy $j$
\begin{itemize}
  \item[$\bullet$]  $(\bol q, \bol p)$ evolves with the noisy
    Hamiltonian dynamics~\eqref{eqn:hamequabruit},
  \item[$\bullet$] each tangent vector $\bol v_i$ evolves with the
    tangent dynamics~\eqref{eqn:nosotan}
  \item[$\bullet$] we use a Gram-Schmidt procedure to turn the
    $\bol v_i$s back into an orthonormal basis: for each $i$,
    \begin{itemize}
    \item we make $\bol v_i$ orthogonal to each $\bol v_{k<i}$: $\bol
      v_i\leftarrow \bol v_i - \sum_{k=1}^{i-1}\bol v_k (\bol v_i
      \cdot \bol v_k)$
    \item we define $s_i(n)=|\bol v_i|$ and then normalize $\bol v_i$:
      $\bol v_i \leftarrow \frac{1}{s_i(n)} \bol v_i$
    \end{itemize}
  \item[$\bullet$] we use the rescaling factors of the
    tangent vectors to compute the weight $w_j(n)=\prod_i
    s_i(n)^{\alpha_i}$
\end{itemize}
\item[\fbox{$2$}] we then compute the average weight 
\begin{equation}
  R(n)=\frac{1}{N_c} \sum _j w_j(n)
\end{equation}
\item[\fbox{$3$}] each copy is then replaced on average by $w_j(n)/R(n)$
  copies. To do so, we pull a random number $\eps_j$ between 0 and
  1. The copy $j$ is replaced by\footnote{$\lfloor x \rfloor$ is the
    largest integer smaller than $x$.}  $\tau=\lfloor
  \eps_j+w_j(n)/R(n)\rfloor$,
  \begin{itemize}
  \item[$\bullet$] if $\tau=0$, we delete the copy
  \item[$\bullet$] if $\tau>1$, we create $\tau-1$ new identical copies,
  \end{itemize}
\item[\fbox{$4$}] at this stage, we have ${N_c}'=\sum_j \lfloor
  \eps_j+w_j(n)/R(n)\rfloor$ copies. On average, we are left with
  ${N_c}$ clones since $\langle{N_c}'\rangle={N_c}$. However, we do not want the number of clones to
  diffuse and thus keep it strictly constant: we kill at
  random ${N_c}'-{N_c}$ clones if ${N_c}'>{N_c}$ and we
  duplicate at random ${N_c}-{N_c}'$ otherwise.
\end{itemize}
The dynamical partition function and free energy are then given by
\begin{equation}
  Z(\alpha,t)=\prod_{n=1}^T R(n)\qquad   \mu(\alpha,t)= \frac 1 t \sum_{n=1}^T \log R(n)
\end{equation}
The distribution of trajectories described by the walkers converge to
the natural distribution of the dynamical systems, weighted with an
extra factor $\exp(\sum_k \alpha_k \lambda_k t)$. Note that as for any
Monte Carlo simulations, metastability can be a problem. The clone
population can indeed get trapped in a metastable ``state'' in the
trajectory space composed of trajectories that are locally favored but
do not globally dominate the average $Z(\alpha)=\langle e^{\sum_k \alpha_k \lambda_k
  t}\rangle $. In such cases, it is better to run several simulations in
parallel and average $\mu(\alpha,t)$ over these simulations rather
than increase the number of clones in one simulation.

\paragraph{Global resampling scheme.} 
An alternative to step $\fbox 2-\fbox 4$ is to resample the whole
population according to the weights $w_j$s. This can be done using
tower sampling~\cite{Krauth}: we construct the cumulative weight
function $C(0)=w_0$ and $C(j\geq 1)=C(j-1)+w_j$. Then we pull ${N_c}$
random numbers between 0 and $C({N_c})$. Every time such a number
falls in the interval $[C(j),C(j+1)]$ we make a new copy of the clone
$j$. This strategy is often used in the Sequential Monte Carlo
literature~\cite{DelMoral2012}. We tried both strategies which gave
similar results. For $\alpha\simeq 0$, the LWD presented above barely
requires any cloning since $\tau\simeq 1$ for all clones. On the other
hand, the global resampling method amounts for $\alpha\simeq 0$ to
choosing $N_c$ random number between $0$ and $N_c$. On average, there
is one number in each interval $[n,n+1]$ but because of fluctuations
many intervals will contain more than one random number and many
intervals will contain no random number. The global resampling then
typically involve many cloning events (typically $N_c/e$ for
$\alpha=0$ and $N_c\gg 1$~\footnote{For large $N_c$, the probability
  that no random number falls in one interval is roughly
  $(1-N_c^{-1})^{N_c}\simeq e^{-1}$ and the average number of empty
  interval is thus $N_c/e$.}). Fluctuations are thus stronger for the
global resampling strategy and the overhead due to the cloning is
typically larger, whence a slower algorithm. The code is, however,
much simpler to write and a large literature exists on how to implement
it efficiently (see~\cite{DelMoral2012} and reference therein).

Another alternative to the LWD, which constructs a canonical measure
over the trajectory space using an inverse temperature $\alpha$, would
be to try and implement a multi-canonical algorithm---a strategy that
was for instance followed in~\cite{Kitajima}. Again, this is left for
future work.

\section{Conclusion}

In this paper, we have presented some more sophisticated applications
of Lyapunov Weighted Dynamics~\cite{Tailleur2007}. In particular, we
have shown that LWD can both be applied to locate atypical
trajectories and to measure free-energy-like quantities in extended
dynamical systems.  This is however not yet the full-blown projects
that one may envisage for the future, such as applications to real
planetary systems, nonlinear waves, and fully developed turbulence.

The method seems indeed a very powerful extension of the usual
sampling of typical trajectories, although one must admit that it
requires, compared to standard Monte Carlo algorithm, much more wisdom
in the choice of parameters such as the number of clones, or the
cloning rate. 

{\bf Acknowledgments}

We wish to thank H. Chat\'e, V. Lecomte, P. Cvitanovic, and F. van
Wijland for useful discussions and A. Solon for sharing several python
macros. This work was partially supported by the grant JAMVIBE of the
Agence Nationale de la Recherche.

\pagebreak

\appendix{}

\section{Conserved noise}
\label{sec:bruitcons}
In this section we show how to derive and implement numerically
Gaussian white noises that conserve the total energy, and show how
they lead to the microcanonical measure in the steady-state. To keep
notations as light as possible we will choose the convention that
repeated indices should be summed over. Furthermore, we write $x_i^2$
instead of $\sum_i x_i^2$.

\subsection{Equations of motion}
For the most generic Hamiltonian $\H$, Hamilton's equations of motion read
\begin{eqnarray}
  \label{eqn:Ham}
  \dot q_i &=& \H_{p_i}\\
      \dot p_i &=& -\H_{q_i}\nonumber
\end{eqnarray}
where we write $\H_{x_i}$ instead of $\ddp{\H}{x_i}$. We then add a
Gaussian white noise $\bol \eta$ and a friction that we adjust to keep
the energy $\H$ constant. The dynamics (\ref{eqn:Ham}) becomes:
\begin{eqnarray}
  \dot q_i &=& {\H}_{p_i}\\
  \dot p_i &=& -{\H}_{q_i} -z {\H}_{p_i} +\sqrt{2 \eps}\eta_i\nonumber
\end{eqnarray}
where $z$ is such that $\dot \H=0$. This implies
\begin{eqnarray}
    \dot \H &=&  {\H}_{q_i} \dot q_i + {\H}_{p_i} \dot p_i={\H}_{q_i} {\H}_{p_i} + {\H}_{p_i} \left(-{\H}_{q_i}-z {\H}_{p_i} +\sqrt{2 \eps} \eta_i \right)\\
  &=&-z  {\H}_{p_i}^2 + \sqrt{2 \eps}  {\H}_{p_i} \eta_i\nonumber
\end{eqnarray}
Finally we set
\begin{equation}
  z=\sqrt{2 \eps}\; \frac{{\H}_{p_i} \eta_i}{{\H}_{p_l}^2}
\end{equation}
so that the equations of motion are given by
\begin{eqnarray}
  \label{eqn:HamNoise}
     \dot q_i &=& {\H}_{p_i}\\ \dot p_i &=& -{\H}_{q_i}-\sqrt{2
       \eps} \frac{ {\H}_{p_k} \eta_k}{
       {\H}_{p_l}^2}\H_{p_i} +\sqrt{2 \eps} \eta_i
     =-{\H}_{q_i}+\sqrt{2 \eps} \underbrace{\left(\delta_{ik} -
         \frac{ {\H}_{p_i} \H_{p_k}}{
           {\H}_{p_l}^2}\right)}_{g_{ik}}\eta_{k}\nonumber
\end{eqnarray}
Let us note that $g$ is the projector onto the energy surface in
the momentum space. Indeed $g$ satisfies
\begin{eqnarray}
  g^2_{ij} &=&g_{ik}g_{kj}
  =\left(\delta_{ik}-\frac{\H_{p_i}\H_{p_k}}{\H_{p_l}^2}\right) \left(\delta_{kj}-\frac{\H_{p_k}\H_{p_j}}{\H_{p_n}^2}\right)
  =\left(\delta_{ij}-2\frac{\H_{p_i}\H_{p_j}}{\H_{p_k}^2}+\frac{\H_{p_i}\H_{p_j}\H_{p_k}^2}{ \H_{p_n}^2  \H_{p_l}^2}\right)\\
  &=&g_{ij}\nonumber
\end{eqnarray}
Furthermore, it is straightforward to check that the vector $\tiny
V=\left(\matrix{ \H_{p_1}\cr \vdots\cr \H_{p_N}}\right) $, which is
normal to the energy surface in the momentum space, is in the kernel
of $g$:
\begin{equation}
  (gV)_i=g_{ij}V_j=\left(\delta_{ij} \H_{p_j}-\frac{\H_{p_i}\H_{p_j}^2}{\H_{p_k}^2}\right)=\H_{p_i}-\H_{p_i}=0
\end{equation}
Hence, $g$ is indeed the projector onto the energy surface.\\
When $\H(\bol q,\bol p)=\frac{\bol p^2}{2}+V(\bol q)$, the dynamics
\eqref{eqn:HamNoise} becomes
\begin{eqnarray}\label{eqn:Hamp2noise}
      \dot q_i &=& p_i\\
      \dot p_i &=& -V_{q_i}- \sqrt{2 \eps} \frac{p_k \eta_k}{ p_l^2} p_i + \sqrt{2 \eps} \eta_i=-V_{q_i} +\sqrt{2 \eps} g_{ij} \eta_j\nonumber
\end{eqnarray}
where $g_{ij}=\delta_{ij}-\frac{p_ip_j}{p^2}$. To keep notations as
simple as possible, we restrict the derivation of the Fokker-Planck
equation to this case, even though it extends to more general
Hamiltonians.

\subsection{Fokker-Planck Equation and steady-state measure}

The noise in~\eqref{eqn:Hamp2noise} is multiplicative and this
Langevin equation is well defined only when a prescription has been
chosen to time-discretize $g(\bol p)$\footnote{The computation leading to
  the conservation of the energy assumes a Stratonovich
  convention~\cite{Oksendal98}.}. Let us now show that under
Stratonovich convention, microcanonical measures are steady-state of
the Fokker-Planck equation. Starting from the Langevin dynamics
\begin{equation}
  \dot \xi_i=h_i(\xi,t)+\sqrt{2 \eps} g_{ij} \Gamma_j
\end{equation}
where $\Gamma_j$ is a Gaussian white noise such that $\langle
\Gamma_j(t) \Gamma_i(t')\rangle = \delta(t-t') \delta_{ij}$, diffusion
and drift coefficients are given by~\cite{Risken96}:
\begin{eqnarray}
  D_i=h_i + \eps g_{kj} \frac{\partial g_{ij}}{\partial
    \xi_k}\qquad\text{and}\qquad D_{ij}= \eps g_{ik}g_{jk}
\end{eqnarray}
which yields for~\eqref{eqn:Hamp2noise}
\begin{eqnarray}
      &D_{q_i}=p_i \qquad \text{and}\qquad D_{p_i}=-V_{q_i}+ \eps g_{p_k p_j}\frac{\partial g_{p_i p_j}}{\partial p_k}\\
      &D_{p_ip_j}= \eps g_{p_ip_k}g_{p_jp_k}
\end{eqnarray}
The Fokker-Planck equation then reads:
\begin{eqnarray}
  \frac{\partial P}{\partial t}&=&\Big[-\frac{\partial}{\partial q_i} p_i + \frac{\partial}{\partial p_i} V_{q_i} + \frac{\partial}{\partial p_i} 
    \eps \big(\frac{\partial}{\partial p_j} g_{p_i p_k} g_{p_jp_k} 
      - g_{p_jp_k} \frac{\partial g_{p_i p_k}}{\partial
        p_j}\big)\Big] P\\
    &=&\Big[-\ddp{}{q_i}p_i+\ddp{}{p_i} V_{q_i}
    +\eps \ddp{}{p_i}\big(g_{p_ip_k}\ddp{}{p_j} g_{p_jp_k}\big)\Big]P
\end{eqnarray}
Using the explicit expression for $g_{p_ip_j}$ and the fact that
$g_{p_ip_j}p_j=0$, we get:
\begin{equation}
\frac{\partial P}{\partial t}=\Big[-\frac{\partial}{\partial q_i} p_i + \frac{\partial}{\partial p_i} V_{q_i} +\eps \frac{\partial }{\partial p_i}\left(\frac{\partial}{\partial p_i}-\frac{p_ip_j}{\sum_r p_r^2}\frac{\partial }{\partial p_j}\right)\Big]P
\end{equation}
Let us now consider the evolution of a distribution $f(\bol q,\bol p)$ that solely
depends on $\H(\bol q,\bol p)$:
\begin{equation}
  \frac{\partial f(\bol q ,\bol p)} {\partial t}= \frac{\partial f(\H)}{\partial t}=-\frac{\partial}{\partial q_i} p_if(\H) 
  + \frac{\partial}{\partial p_i} V_{q_i}f(\H) 
  +\eps\frac{\partial }{\partial p_i}\left[\frac{\partial}{\partial p_i}f(\H)-\frac{p_ip_j}{p^2}\frac{\partial }{\partial p_j}f(\H)\right]
\end{equation}
Using $\frac{\partial f(\H)}{\partial x_i}=f'(\H)\H_{x_i}$, we get:
\begin{equation}
  \frac{\partial f[\H(\bol q,\bol p)]}{\partial t}=
  -f'(\H)\left[\frac{\partial \H}{\partial q_i} p_i
    -\frac{\partial \H}{\partial p_i} V_{q_i} \right] 
  +\eps \frac{\partial }{\partial p_i} 
  \left[f'(\H)\left(\frac{\partial \H}{\partial p_i}
      -\frac{p_ip_j}{p^2} \frac{\partial \H}{\partial p_j}\right)\right]=0
\end{equation}

We have thus shown how to construct a Langevin equation that conserves
the energy and yield a uniform measure onto the energy surface. Let us
further notes that the numerical implementation
of~\eqref{eqn:HamNoise} could be rather difficult, involving a
cumbersome multiplicative noise. Fortunately, there is a very simple
way around this, exact at the order $\sqrt {\eps \dt}$ but which
conserves the energy \textit{exactly}. We pull a random vector $\bol
\eta$ on a sphere of dimension $N$ and radius $\sqrt{2 \eps \dt}$,
which we add to $\bol p$. We then renormalise $\bol p$ to keep its
norm constant. This amounts to
\begin{equation}
  \bol p \to \frac{|\bol p|}{|\bol p+ \sqrt{2 \eps \dt}\, \bol \eta|} \left( \bol p + \sqrt{2 \eps \dt}\, \bol \eta\right)
\end{equation}
At first order in $\sqrt {2 \eps \dt}$, this reads:
\begin{equation}
  \bol p \to \bol p + \sqrt{2 \eps \dt} \left[\bol \eta - \bol p\, \frac{\bol p \cdot \bol \eta}{|\bol p|²}\right]
\end{equation}

If one further wishes to conserve a total impulsion $\sum_i p_i=0$, we
simply replace $\eta_i$ by $\eta_i - \frac 1 N \sum_j \eta_j$ in the
above procedure. Last, let us note that despite our manipulations,
these noises remain Gaussian, as linear combinations of Gaussian noises.

~


\begin{thebibliography}{99}
\bibitem{predrag} The exercise was suggested to us by Predrag Cvitanovic.
\end{thebibliography}
\end{document}